\documentclass[aps,preprint,showpacs]{revtex4}

\usepackage{epsfig}
\usepackage{color}
\usepackage{graphicx}

\begin{document}

\title{When is a surface foam-phobic or foam-philic?}

\author{Miguel A. C. Teixeira$^*$}
\affiliation{Department of Meteorology, University of Reading \\
Earley Gate, PO Box 243, Reading RG6 6BB, United Kingdom}
\email{m.a.teixeira@reading.ac.uk}

\author{Steve Arscott$^\dagger$}
\affiliation{Institut d'Electronique, de Micro\'{e}lectronique et de 
Nanotechnologie (IEMN) \\ CNRS UMR8520, The University of Lille \\
Cit\'{e} Scientifique, Avenue Poincar\'{e}, 59652 Villeneuve d'Ascq, France}
\email{steve.arscott@iemn.univ-lille1.fr}

\author{Simon J. Cox$^\ddagger$}
\affiliation{Department of Mathematics, Aberystwyth University \\ 
Aberystwyth, Ceredigion, SY23 3BZ, United Kingdom}
\email{foams@aber.ac.uk}

\author{Paulo I. C. Teixeira$^{\S}$}
\affiliation{ISEL -- Instituto Superior de Engenharia de Lisboa,
Instituto Polit\'{e}cnico de Lisboa \\
Rua Conselheiro Em\'{\i}dio Navarro 1, 1959-007 Lisbon, Portugal}
\affiliation{Centro de F\'isica Te\'orica e Computacional, Faculdade de
Ci\^{e}ncias da Universidade de Lisboa \\
Campo Grande, Edif\'{\i}cio C8, 1749-016 Lisbon, Portugal}
\email{piteixeira@fc.ul.pt}

\date{15 September 2017}

\begin{abstract}
By integrating the Young-Laplace equation, including the effects of gravity,
we have calculated the equilibrium shape of the two-dimensional Plateau borders 
along which a vertical soap film contacts two flat, horizontal solid substrates 
of given wettability. We show that the Plateau borders, where most of a foam's 
liquid resides, can only exist if the values of the Bond number ${\rm Bo}$ 
and of the liquid contact angle $\theta_c$ lie within certain domains in 
$(\theta_c,{\rm Bo})$ space: under these conditions the substrate is 
foam-philic. For values outside these domains, the substrate cannot support 
a soap film and is foam-phobic. In other words, on a substrate of a given 
wettability, only Plateau borders of a certain range of sizes can form. For 
given $(\theta_c,{\rm Bo})$, the top Plateau border can never have greater 
width or cross-sectional area than the bottom one. Moreover, the top Plateau 
border cannot exist in a steady state for contact angles above 90$^\circ$. 
Our conclusions are validated by comparison with both experimental and 
numerical (Surface Evolver) data. We conjecture that these results will hold,
with slight modifications, for non-planar soap films and bubbles. 
Our results are also relevant to the motion of bubbles and foams in channels, 
where the friction force of the substrate on the Plateau borders plays an 
important role.


\end{abstract}

\maketitle

\section{Introduction}

\label{sect-intro}

The wetting of a solid by a liquid -- where the liquid will spread into a 
sheet or break up into droplets when placed onto the solid -- is ubiquitous 
in nature as well as having practical importance in industry 
\cite{Rowlinson:1982}. Wetting behaviour can be conveniently described 
in terms of the contact angle $\theta_c$ at which the liquid-vapour interface 
meets the solid-liquid interface: if $\theta_c=0$ the liquid is said to 
completely (or perfectly) wet the solid, whereas if $0<\theta_c\le\pi/2$ 
wetting is only partial. Contact angles greater than $\pi/2$ correspond 
to drying (or de-wetting) of the solid by the liquid. 
If, as is often the case in practice, 
the liquid is water-based, a surface that is wetted ($0\le\theta_c\le\pi/2$) 
is called hydrophilic, and one that is not ($\pi/2<\theta_c\le\pi$) hydrophobic.
If $\theta_c$ is greater than about $5\pi/6\,(150^\circ)$ there is only a very 
small area of contact between liquid and solid: essentially the liquid forms 
an almost spherical droplet that may, e.g., under the effect of gravity, roll 
off the solid, which is then termed superhydrophobic. Superhydrophobicity is 
a topic of much current research: see, e.g., 
\cite{Gao:2009a,Gao:2009b,Shirtcliffe:2010} for reviews.

\begin{figure}
\begin{center}
\psfig{figure=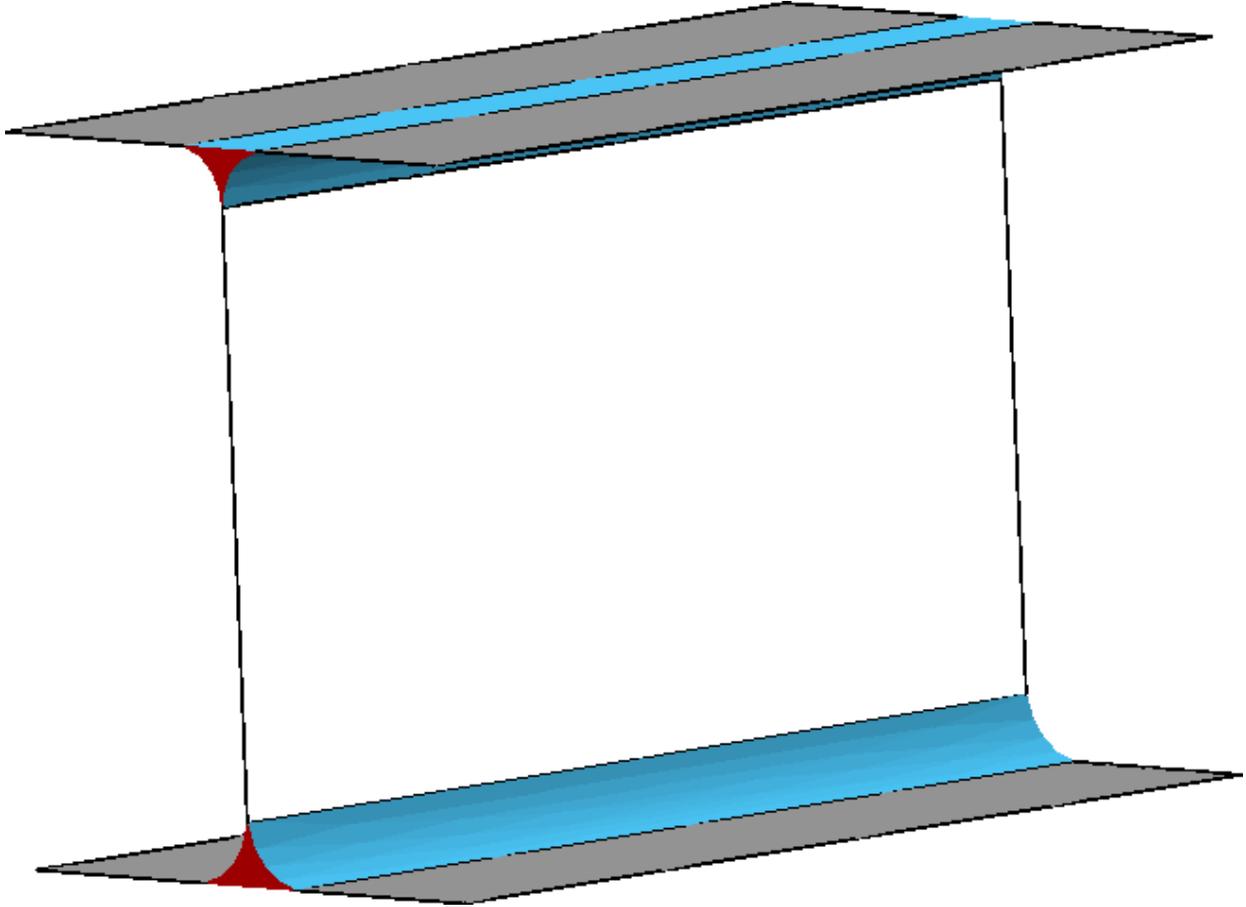,width=\textwidth}
\end{center}
\caption{Surface-Evolver-generated oblique view of a soap film spanning 
the gap between two parallel walls. The film (transparent) meets the walls 
(grey) at surface Plateau borders (blue). Each surface Plateau border is 
bounded by the solid wall and by two curved liquid-vapour interfaces. If 
the film is planar then the Plateau borders have uniform cross-section 
(red) along a direction parallel to both the film and the walls, and is 
thus effectively 2D.}
\label{figwallPB}
\end{figure}

In confined foams, which include most real-life foams, there are, in addition 
to the usual bulk Plateau borders at which three (or more, in a wet foam) soap 
films meet, Plateau borders where the films meet the confining walls. These 
surface Plateau borders are bounded by the wall and (in fairly dry foams) 
by two curved liquid-vapour interfaces, see figure \ref{figwallPB}. 
These interfaces will, of course, meet the 
wall at the liquid contact angle $\theta_c$. The question thus arises: what 
is the shape of a Plateau border of a given size (i.e., volume in 3D, area in 
2D) on a surface of a given wettability? Or, in other words, when is a solid 
surface capable of supporting a foam, i.e. is the surface `foam-phobic' or 
`foam-philic'? This is of paramount importance for assessing, e.g., the 
effectiveness of firefighting foams on different substrates, or the 
adequacy of containers for certain foamy foodstuffs.

In an earlier paper, we calculated the shape of a 2D surface Plateau border 
around a bubble sitting on a perfectly-wetting substrate in zero gravity 
\cite{Teixeira:2005}. This was later extended to 3D and a (fairly small but)
finite contact angle \cite{Teixeira:2007}, and to include the effect of gravity
\cite{Teixeira:2009}. More recently, we calculated the equilibrium shape of the
axially symmetric meniscus along which a bubble contacts a flat {\it liquid}
surface \cite{Teixeira:2015}. Here we return to solid surfaces of variable 
wettability, but consider first the simpler case of a planar film spanning a 
gap between two parallel, flat substrates (a rectangular slit). The film and 
its associated surface Plateau borders are thus effectively 2D (slab-symmetric):
see figure \ref{figwallPB}.

Most of a foam's liquid is contained within the network of Plateau borders. 
It is clear from the curvature of the Plateau border interfaces that the 
liquid is at lower pressure than the gas in the bubbles; indeed the interfacial 
curvature is set by this pressure difference and so, at the same height in a 
foam, the liquid in the bulk and surface Plateau borders will have the same 
pressure. As a consequence of their different shapes, however, the surface 
Plateau border (with one of its interfaces in contact with the planar wall) 
will have greater volume \cite{Cantat:2013a}. Thus, per unit length, the 
surface Plateau borders carry a disproportionately large amount of a foam's 
liquid, and therefore understanding their shape and stability is important. 
Moreover, as a foam moves, it is the surface Plateau borders that drag along 
the substrate and set one of the important time-scales for foam dynamics 
\cite{Cantat:2013b}.

This paper is organised as follows: in section \ref{sect-exper} we describe
our experimental method for measuring Plateau border shapes. These shapes can 
be found analytically for arbitrary gravity and liquid contact angle, which we 
do by solving the Young-Laplace equation in section \ref{sect-theor}. We then 
derive the ranges of parameters for which such Plateau borders may exist,
which is a necessary condition to form a foam on a surface of given wettability.
An alternative method to find Plateau border shapes from numerical energy 
minimisation, using the Surface Evolver software, is described in section 
\ref{sect-num}. Then in section \ref{sect-res} we compare the predictions 
of our analytical solution with experimental results, as well as with the 
fully numerical Surface Evolver solution. Finally, we conclude in section 
\ref{sect-disc}.

\section{Experimental procedure}

\label{sect-exper}

The film-surface wetting experiments were performed in a dust-free, controlled 
environment, using a class ISO 5/7 cleanroom, which ensures that the 
temperature ($T$) and relative humidity ($RH$) remain within the following 
ranges: $T = 20\pm 0.5^{\circ}$C and $RH=45\pm 2\%$.
The data were gathered using a contact angle meter (GBX Scientific Instruments,
France). A commercially-available surfactant solution (Pustefix, Germany) was 
employed to generate stable soap films for the experiments. This solution 
is a mixture of water, glycerol, and an organosulfate. The surface tension 
and the density of the solution have previously been measured to be 
$28.2\pm 0.3~{\rm mJ m}^{-2}$ and $997.8~{\rm kg m}^{-3}$ \cite{Arscott:2013}.

\begin{figure}
\begin{center}
\psfig{figure=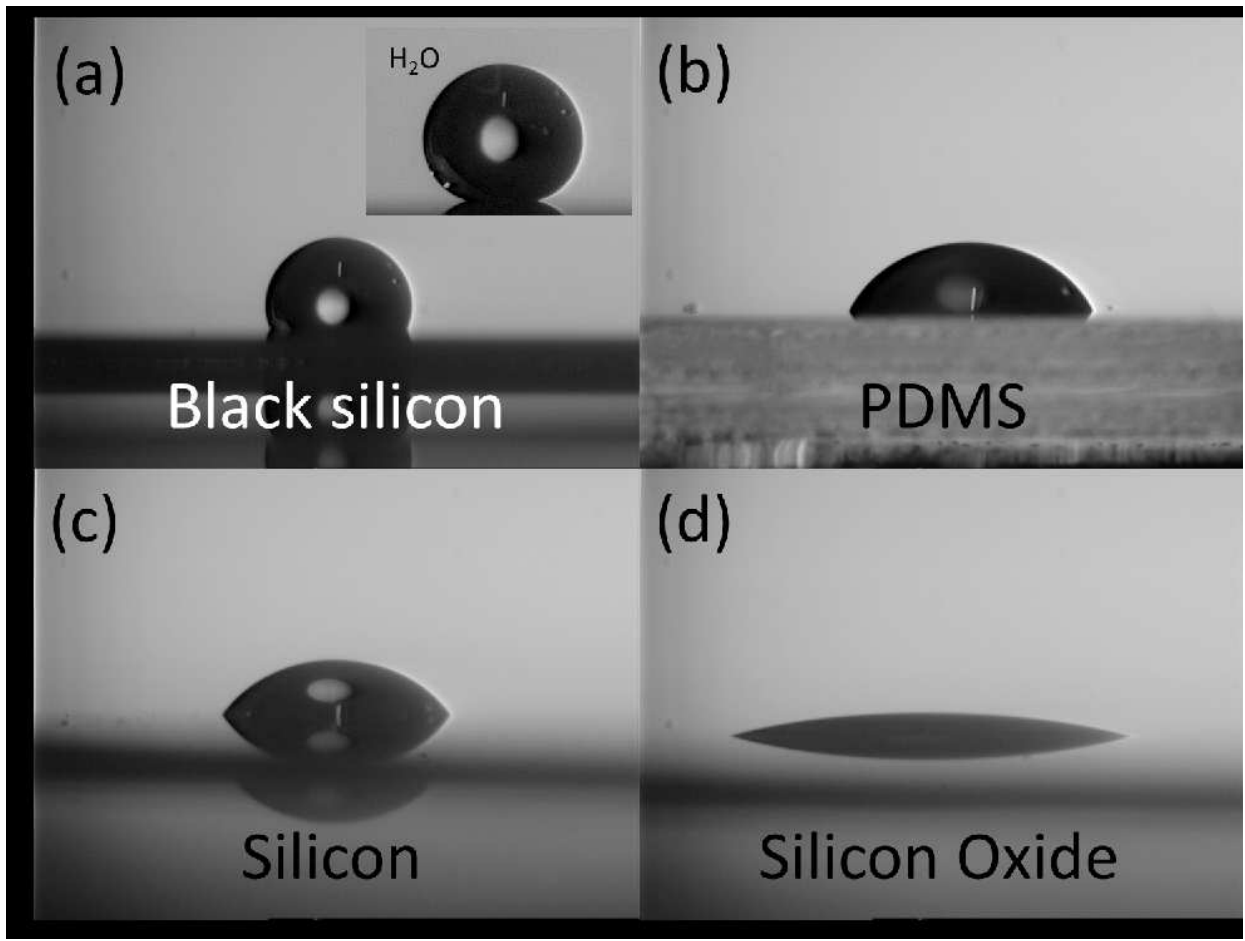,width=\textwidth}
\end{center}
\caption{Droplets of the bubble solution on four of the five different 
surfaces used in the experimental part of the study: (a) teflonised black 
silicon; (b) PDMS elastomer; (c) teflonised polished silicon; and (d) silicon 
oxide. The inset to panel (a) shows a water droplet resting on a teflonised 
black silicon surface. The droplet base diameters in (a) to (d) are 1.9~mm, 
3.4~mm, 3.2~mm and 5.6~mm. The diameter of the droplet in the inset to (a) 
is 2.6~mm.}
\label{figdrops}
\end{figure}

Five different solid surfaces were used in the experiments, the properties of 
which are collected in Table \ref{table1}. These five surfaces were prepared 
to ensure a range of wetting properties (with respect to the surfactant 
solution) from hydrophilic (low contact angle) to hydrophobic 
(high contact angle). A hydrophilic surface was prepared by 
chemically oxidising a commercial p-type ($5-10~\Omega{\rm cm}$) 
polished silicon wafer (Siltronix, France) in $65\%$ nitric acid, thus 
creating a thin silicon oxide layer having a thickness of approximately 
1~nm \cite{Arscott:2016} (surface 1). Three intermediate wetting surfaces were 
fabricated from a commercial polished silicon wafer (roughness $<1$ nm) 
coated with a thin amorphous fluorocarbon (FC) layer \cite{Zhuang:2005} -- 
referred to here as `teflonised polished silicon' (surface 2); a 1~mm thick 
polydimethylsiloxane `PDMS' elastomer block (1:10 PDMS Sylgard 184 Dow Corning)
moulded in a dish (surface 3); and a `teflonised rough silicon' surface 
made by depositing the thin amorphous FC on the unpolished rear side of a 
commercial silicon wafer of roughness $\approx 1~\mu{\rm m}$ (surface 4). 
A hydrophobic surface was prepared by coating `black silicon', prepared using 
a Bosch\textregistered{} process etch under certain plasma conditions 
\cite{Stubenrauch:2006}, with an FC layer -- this is referred to here as 
`teflonised black silicon' (surface 5). 
The FC layer was deposited by exposure of the surfaces (both silicon and 
black silicon) to a C$_4$F$_8$ plasma (Surface Technology Systems Ltd, UK), 
which resulted in the deposition of a thin (a few tens of nanometres) 
film of amorphous fluoropolymer on the surface of both the silicon and 
the black silicon. The teflonised black silicon was verified to be 
superhydrophobic: its wetting contact angle to water droplets was measured 
to be $154.5\pm 2.4^{\circ}$ with near-zero contact angle hysteresis. The 
wetting contact angle of the surfactant solution was measured on each surface 
using the contact angle meter (see table \ref{table1}) and the results are 
consistent with previous measurements \cite{Arscott:2013}. Figure 
\ref{figdrops} shows photographs of droplets of the surfactant solution 
on four of the five different surfaces described above.

\begin{figure}
\begin{center}
\psfig{figure=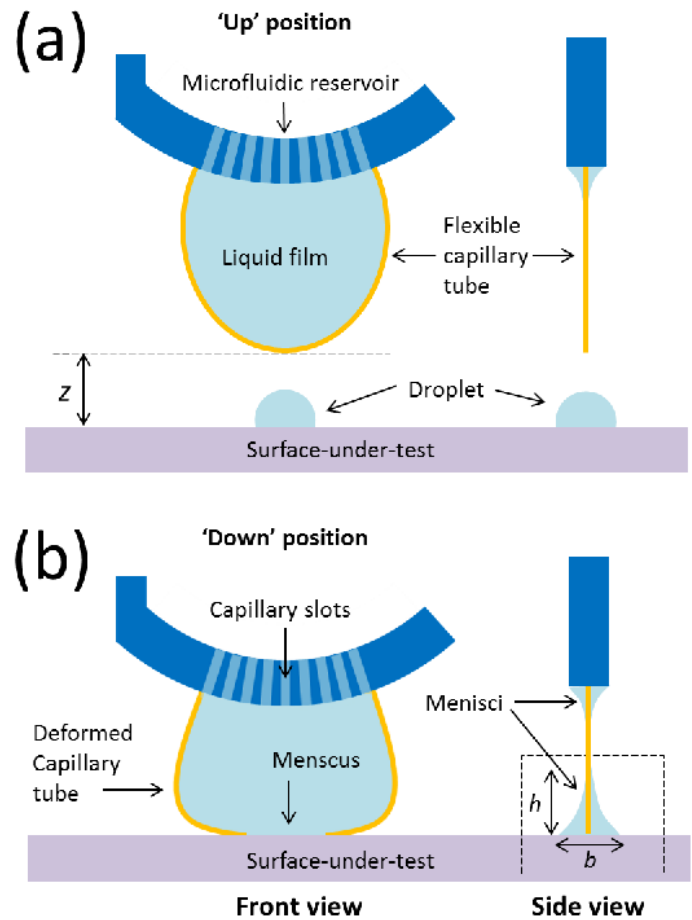,width=10cm}
\end{center}
\caption{Schematic diagram showing the experimental setup -- front view and 
side view. The in-house microfluidic tool is in (a) the `up position', and 
(b) the `down position'. The tool consists of a microfluidic reservoir 
(dark blue) and a deformable loop (gold) holding the liquid film 
(light blue). The tool is placed inside the contact angle meter. 
The dashed box indicates the photograph shown in figure \ref{figmenisci}.}
\label{figsetup}
\end{figure}

A schematic diagram of the experimental setup and the working principle is shown
in figure \ref{figsetup}. It contains an in-house microfluidic tool which has 
been created specifically for the experiments. The tool incorporates two main 
elements: a microfluidic reservoir and a deformable ring, made of a loop of 
capillary tube. The role of the microfluidic reservoir is to 
increase the lifetime of the liquid film sufficiently to allow the formation 
of a stable Plateau border (see figure \ref{figsetup}b).
The lifetime of the liquid film was approximately 30~s in the current 
setup, which was sufficient to create a stable Plateau border and photograph 
it. An unwanted side-effect is that there is gravity-driven drainage from the 
reservoir towards the Plateau border, so that its volume increases 
during the experiment. 

The role of the deformable loop is threefold: (i) to support 
a stable liquid film connected to the reservoir; (ii) to be thin enough so 
as not to perturb the Plateau border shape, e.g., thickness of loop 
very much less than the Plateau border dimensions $h$ and $b$ 
(see figure \ref{figsetup}b); and (iii) the loop 
should be deformable to enable the formation of a long, stable Plateau border
across the surface. Indeed, this deformability -- leading to a long, 
voluminous Plateau border -- combats the effect of drainage from the reservoir. 
The radius $R$ of the deformable loop in the current setup is 
$\approx 1~{\rm cm}$. 

We bring the tool containing the liquid 
film (figure \ref{figsetup}a -- `up' position) carefully into contact with 
a small droplet resting on the specific surface under test (figure 
\ref{figsetup}b -- `down' position) in the contact angle meter. 
Upon contact, and allowing the loop to be slightly deformed as shown 
in figure \ref{figsetup}b, a long stable Plateau border is formed along the 
surface, over a length of about 1~cm, which can be photographed 
(side view in figure \ref{figsetup}b) using the contact angle meter.

\begin{figure}
\begin{center}
\psfig{figure=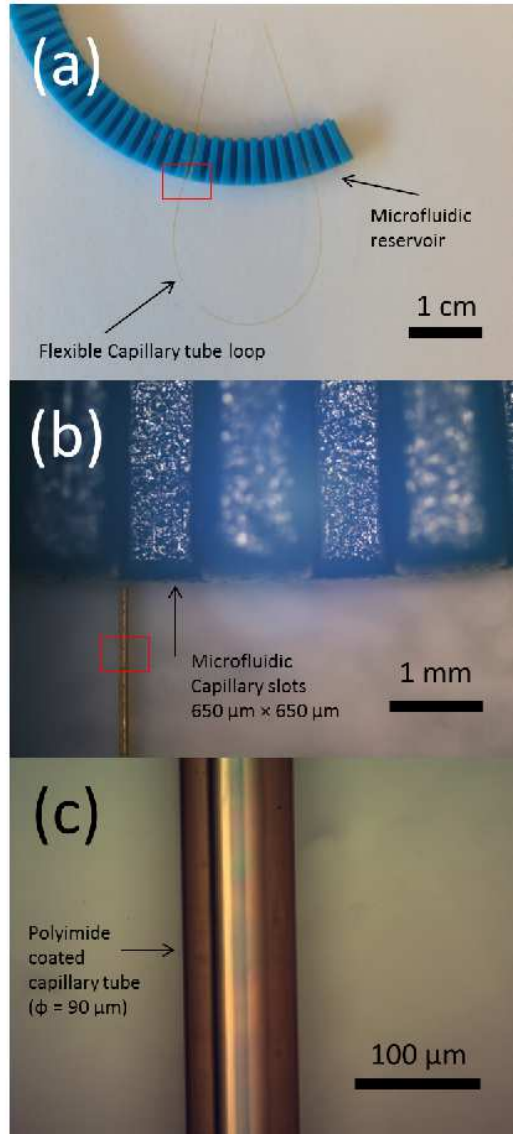,width=7cm}
\end{center}
\caption{Photographs of the parts of the microfluidic tool. 
(a) The microfluidic reservoir (blue) containing the capillary slots and the 
capillary tube which forms a deformable loop; (b) Zoom of the microfluidic 
capillary slots made of plastic (ABS); and (c) zoom of the flexible 
polyimide-coated, fused silica capillary tube (outside diameter 
$90~\mu{\rm m}$). The red boxes indicate the zoom regions.}
\label{figmicrof}
\end{figure}

Figure \ref{figmicrof} shows the practical components of the microfluidic tool. 
The reservoir is contained within capillary slots with a width and depth of 
$\approx 650~\mu{\rm m}$ and a length of 6~mm (there are 14 on the tool, 
holding a total liquid volume of about $35~\mu$l) made of ABS plastic. 
The loop which supports the liquid film is made of polyimide-coated capillary 
tubing (Molex, USA) having an outside diameter of $90~\mu{\rm m}$.

In the current setup it is very difficult to have the deformed loop perfectly 
perpendicular to the camera -- this is visible at the top of the Plateau 
borders in the photographs (see figure \ref{figmenisci}). Moreover, as the loop
is deliberately not rigid, the attached film can vibrate. A rough estimation 
of the soap film vibration frequency $f$ (first mode) of a circular loop can 
be made by using $f=(1/2\pi)\sqrt{\gamma/\pi\rho R^2 t}$, where $\gamma$, 
$\rho$ and $t$ are the soap film surface tension, density and thickness, and 
$R$ is the loop radius. Taking $t$ to be $1~\mu{\rm m}$ and the values given 
in the text for the other quantities, one can estimate $f\approx 50~{\rm Hz}$. 
In some cases, the amplitude of such oscillations can be of the order of 
millimetres \cite{Drenckhan:2008}. This effect can contribute to blur in the
photographs (low lighting -- longer shutter times) at the top of the Plateau 
border. Another source of experimental error is non-perfect surfaces, which 
is apparent from the fact that the contact angles are not always equal on the 
left and on the right of the Plateau borders. This is inevitable despite care 
(e.g., working in a cleanroom in this case -- surface preparation, storage 
and measurements): there are defects and contamination that cause wetting to 
be asymmetrical.

Finally, note that this setup only allows us to measure the Plateau border
at the bottom substrate, not at the top one.

\section{Analytical theory}

\label{sect-theor}

\begin{figure}
\begin{center}
\psfig{figure=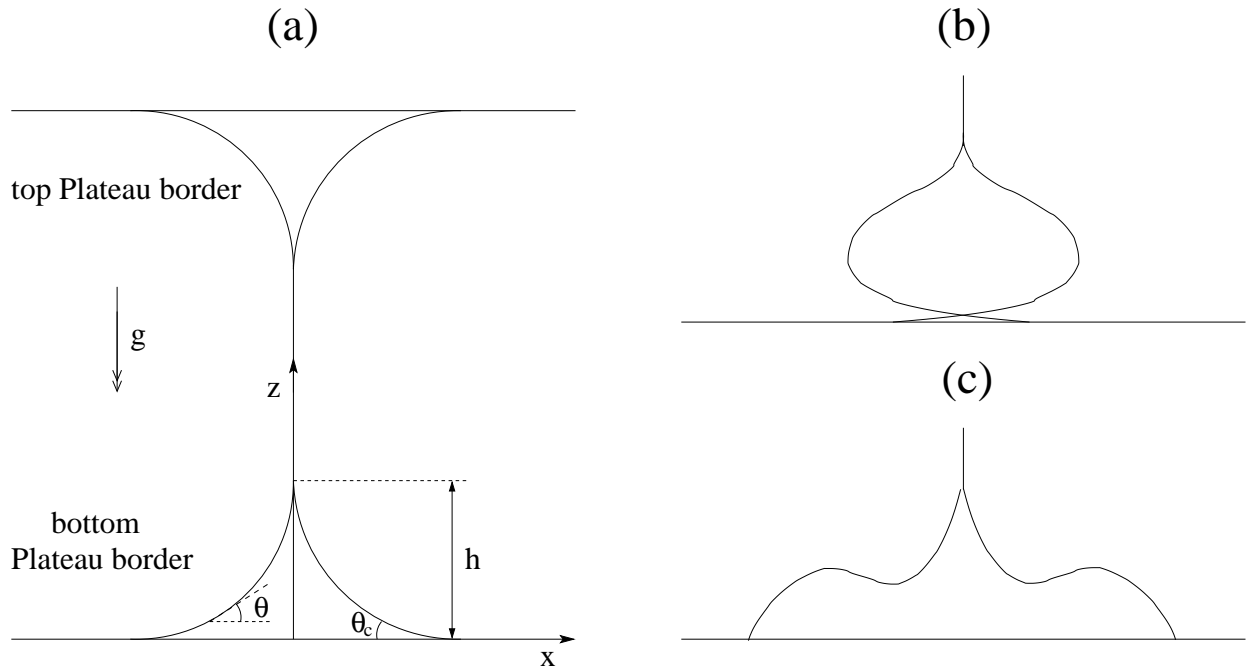,width=\textwidth}
\end{center}
\caption{(a) Sketch of a slab-symmetric soap film spanning the gap between two
flat horizontal substrates and the associated surface Plateau borders: $z$ is 
the height, $x$ is the distance from the film (the $z$-axis) to the Plateau 
border surface, $h$ is the Plateau border height, $\theta$ is the Plateau 
border inclination, and $\theta_c$ is the liquid contact angle at the 
substrate, located at $z=0$. The gravitational acceleration is $g$. (b) Sketch 
of an unphysical surface Plateau border in the lower forbidden domain of figure 
\ref{fig1}a. (c) Sketch of an unphysical surface Plateau border in the upper 
forbidden domain of figure \ref{fig1}a.}
\label{fig0}
\end{figure}

The Young-Laplace law for the 2D (i.e., slab-symmetric) liquid surfaces 
bounding a Plateau border at a flat substrate (see figure \ref{fig0}a) 
can be written 
\cite{Isenberg:1978}:
\begin{equation}
\left[ 1 + \left( \frac{d x}{d z} \right)^2 \right]^{-3/2}
\frac{d^2 x}{d z^2} = - \frac{\Delta p}{\gamma},
\label{laplace2d}
\end{equation}
where $z$ is height measured from the substrate, $x$ is the 
distance measured horizontally from the plane of symmetry (the plane of 
the 2D film), $\Delta p(z)$ is the pressure difference across the liquid 
surface at each height, and $\gamma$ is the surface tension of the liquid.

Our aim is to solve equation (\ref{laplace2d}) for one of the surfaces bounding 
each of the top and bottom Plateau borders of a 2D vertical film spanning 
the gap between two flat, horizontal substrates. Naturally, the other Plateau 
border surface is mirror-symmetric with respect to $x=0$. The Plateau borders 
are slab-symmetric and in hydrostatic equilibrium. It is convenient to choose a
definition of $\Delta p$ that, by default, makes $x$ predominantly positive 
in the most common situations. This is $\Delta p=p_b-p_a$, where $p_b$ is the 
pressure inside the Plateau border (i.e., within the liquid) and $p_a$ is the 
atmospheric pressure outside the Plateau border (assumed to be constant). 

We shall start by considering the bottom Plateau border. Since $p_b$ is assumed
to be in hydrostatic equilibrium, we have
\begin{equation}
\Delta p= p_b - p_a = p_{b0} - p_a - \rho g z,
\label{hydrostatic}
\end{equation}
where $p_{b0}$ is the pressure inside the Plateau border at the substrate 
($z=0$), $g$ is the gravitational acceleration, and $\rho$ is the density 
of the liquid inside the Plateau border (in our case water). 

Additionally, we introduce the convenient change of variables given by
\begin{equation}
\frac{d x}{d z} = - \cot \theta \quad \Rightarrow
\quad \frac{d^2 x}{d z^2} = {\rm cosec}^2 \theta 
\frac{d \theta}{d z},
\label{changevar}
\end{equation}
where $\theta$ is the angle of inclination of the film surface (see figure 
\ref{fig0}a). Using equations (\ref{hydrostatic}) and (\ref{changevar}), 
equation (\ref{laplace2d}) becomes
\begin{equation}
\sin \theta \frac{d \theta}{d z} = \frac{p_a - p_{b0}}{\gamma}
+ \frac{\rho g z}{\gamma}.
\label{eqangle}
\end{equation}
This equation can be straightforwardly solved for $\theta$, yielding
\begin{equation}
\cos \theta(z) = \cos \theta_c -\frac{p_a - p_{b0}}{\gamma} z - \frac{\rho
g}{2 \gamma} z^2,
\label{soltheta}
\end{equation}
where the integration has been carried out from the base of the Plateau border,
$z=0$, where $\theta=\theta_c$, to a generic height $z$. By definition, 
$\theta_c$ is the contact angle of the liquid with the underlying solid 
substrate, and varies in the interval $0< \theta_c < \pi$. If equation 
(\ref{eqangle}) is instead integrated from $z=0$ to the top of the Plateau 
border $z=h$, where it is assumed that the film is vertical (i.e., 
$\cos\theta=0$, so $\theta=\pi/2$), this provides a definition for the 
pressure term on the right-hand side of the solution, equation 
(\ref{soltheta}), which allows us to eliminate this term:
\begin{equation}
\frac{p_a - p_{b0}}{\gamma}= \frac{1}{h} \cos \theta_c - \frac{\rho g h}{2 
\gamma}.
\label{defpres}
\end{equation}
Equation (\ref{soltheta}) can now be expressed entirely in terms of $z$, $h$ 
and $\theta_c$, as follows:
\begin{equation}
\cos \theta(z) = \cos \theta_c \left(1 - \frac{z}{h} \right) + \frac{\rho g z}
{2 \gamma} ( h - z).
\label{soltheta2}
\end{equation}
This equation can be written more simply if $z$ is made dimensionless
by scaling it by $h$, $z^\prime=z/h$, and a Bond number is defined as 
${\rm Bo}=\rho g h^2/\gamma$. In terms of these quantities, equation 
(\ref{soltheta2}) can be rewritten as 
\begin{equation}
\cos\theta(z^\prime) = (1-z^\prime)\left(\cos\theta_c +\frac{Bo}{2} z^\prime\right).
\label{soltheta3}
\end{equation}

To obtain $x$ as a function of $z$, we now go back to the definition 
of $dx/dz$. Further defining $x^\prime=x/h$, it follows that
\begin{equation}
\frac{d x^\prime}{d z^\prime}=\frac{d x}{d z} = -\cot \theta = 
-\frac{\cos \theta}{\sqrt{ 1 - \cos^2 \theta }}.
\label{slopedef}
\end{equation}
Using equation (\ref{soltheta3}), equation (\ref{slopedef}) can be rewritten as
\begin{equation}
\frac{dx^\prime}{d z^\prime}=-\frac{(1-z^\prime)\left(\cos\theta_c
+\frac{\rm Bo}{2}z^\prime\right)}{\left[1-(1-z^\prime)^2
\left(\cos\theta_c+\frac{Bo}{2} z^\prime\right)^2 \right]^{1/2}}.
\label{eqdxdz}
\end{equation}
Noting again that at the top of the Plateau border
$x^\prime(z=h)=x^\prime(z^\prime=1)=0$, equation (\ref{eqdxdz}) 
can be integrated between a generic $z^\prime$ and $z^\prime=1$, yielding
\begin{equation}
x^\prime(z^\prime) = \int_{z^\prime}^{1} \frac{(1 - z^{\prime\prime}) 
\left( \cos \theta_c 
+ \frac{\rm Bo}{2} z^{\prime\prime}\right)}{\left[ 1 - ( 1 - 
z^{\prime\prime})^2 \left( \cos \theta_c + \frac{\rm Bo}{2}
z^{\prime\prime} \right)^2 \right]^{1/2}} 
dz^{\prime\prime} \qquad\mbox{(bottom PB)}.
\label{defxfinb}
\end{equation}
This equation gives the shape of the right-hand surface bounding the bottom
Plateau border (between $z^\prime=0$ and $z^\prime=1$). The shape of the top 
Plateau border is then immediately obtained by reversing the sign of $g$ in
equation (\ref{hydrostatic}) and following through the above derivation, 
with the result 
\begin{equation}
x^\prime(z^\prime) = \int_{z^\prime}^{1} \frac{(1 - z^{\prime\prime}) 
\left( \cos \theta_c 
- \frac{\rm Bo}{2} z^{\prime\prime}\right)}{\left[ 1 - ( 1 - 
z^{\prime\prime})^2 \left( \cos \theta_c - \frac{\rm Bo}{2}
z^{\prime\prime} \right)^2 \right]^{1/2}} 
dz^{\prime\prime} \qquad\mbox{(top PB)}.
\label{defxfint}
\end{equation}
Note that, while the meaning of $x^\prime$ remains unchanged, in this case 
$z^\prime$ is a dimensionless height measured as positive downwards from the 
top substrate.

Another relevant quantity is the area delimited by the Plateau border
cross-section. This is defined as
\begin{equation}
A=2 \int_0^h x \, dz = 2 \left[ z x \right]_0^h 
- 2 \int_0^h  z \frac{d x}{dz} \, dz= - 2 \int_0^h z \frac{d x}
{dz} \, dz,
\label{areadef}
\end{equation}
where the second equality results from integrating by parts,  and the 
third equality follows from the fact that at the top of the Plateau border 
$x(z=h)=0$.
The factor of 2 in equation (\ref{areadef}) 
accounts for the fact that the Plateau border surfaces are symmetric. Defining 
a dimensionless area as $A^\prime=A/h^2$, this is given, from equation 
(\ref{areadef}), by
\begin{equation}
A^\prime = - 2 \int_0^1 z^\prime \, \frac{d x^\prime}{d z^\prime} \, dz^\prime.
\label{ndadef}
\end{equation}
Using equation (\ref{eqdxdz}), equation (\ref{ndadef}) for the bottom Plateau 
border can be written explicitly as
\begin{equation}
A^\prime=2\int_0^1 \frac{z^\prime (1-z^\prime) \left(\cos\theta_c+\frac{\rm Bo}
{2} z^\prime \right)}{\left[ 1 - ( 1 - z^\prime)^2 \left(\cos\theta_c
+\frac{\rm Bo}{2}z^\prime \right)^2 \right]^{1/2}} dz^\prime
\qquad\mbox{(bottom PB)},
\label{adeffinalb}
\end{equation}
and the equivalent result for the top Plateau border is 
\begin{equation}
A^\prime=2\int_0^1 \frac{z^\prime (1-z^\prime) \left(\cos\theta_c-\frac{\rm Bo}
{2} z^\prime \right)}{\left[ 1 - ( 1 - z^\prime)^2 \left(\cos\theta_c
-\frac{\rm Bo}{2}z^\prime \right)^2 \right]^{1/2}} dz^\prime
\qquad\mbox{(top PB)}.
\label{adeffinalt}
\end{equation}
These results will be compared with experimental and simulated Plateau border 
shapes and areas in section \ref{sect-res}.

\section{Numerical method}

\label{sect-num}

We predict the shape of both the bottom and top Plateau borders numerically 
using Brakke's Surface Evolver \cite{brakke92}. We use cgs units throughout: 
the substrate separation in the $z$ direction is 2~cm (which is arbitrary 
provided the top and bottom Plateau borders do not touch), 
the value of gravity is taken to be $981~{\rm cm/s}^2$, liquid density is 
$1~{\rm g/cm}^3$; then Plateau border areas are measured in ${\rm cm}^2$ and 
surface tensions in mN (note that this is a 2D `line' tension).

The simulation consists of just one half of the domain, by symmetry (see 
figure \ref{fig0}a), using just three fluid interfaces: one for the top 
Plateau border, one for the bottom Plateau border, and one for the vertical 
film joining them. All three have surface tension  $\gamma = 28.2~{\rm mN}$, 
since the vertical film is one half of the physical double interface. To 
specify the contact angle $\theta_c$ at which the Plateau borders meet the 
substrates, we insert a further wetting film along the substrates, outside 
the Plateau borders, with tension $\gamma_{\rm wall} = \gamma \cos \theta_c$. 

Each Plateau border has fixed area and the two areas can be varied 
independently. We used a top Plateau border half-area of $0.005~{\rm cm}^2$ 
throughout, and increased the bottom Plateau border area to the required 
value (of up to about $0.3~{\rm cm}^2$) from an initial half-area of 
$0.020~{\rm cm}^2$ to explore different Bond numbers. Similarly, the contact 
angle $\theta_c$ is increased from zero in steps of one degree to allow all 
values up to $180^\circ$ to be explored.

To allow the Plateau border surfaces to curve, each interface is discretised 
into $N$ short straight segments; we expect a better representation of
the interface at higher $N$, and illustrate the convergence to the
analytic solution with increasing $N$ in figure \ref{fig8}.
The Surface Evolver is used to minimise the free 
energy of the system, i.e., the product of length and surface tension of the 
interfaces, subject to the fixed Plateau border areas. We evaluate the Hessian 
of energy frequently to ensure that the arrangement of films is a stable one 
\cite{brakke96}.

The results of the simulation include Plateau border heights and widths, 
and the three interface lengths for different contact angles and Plateau 
border areas. They are compared with our theoretical predictions in the
next section.

\section{Results and discussion}

\label{sect-res}

In this section we compare thoretical, simulated and experimental results 
for Plateau border shapes. We first consider the shapes without gravity 
(\ref{sect-res-nograv}). When gravity is included, that is at finite Bond 
number, we consider the bottom and top Plateau borders separately 
(\ref{sect-res-bot} and \ref{sect-res-top}, respectively). For the 
bottom Plateau border, the experiments generate a variation in the Bond 
number by varying the size of the Plateau border, i.e., its liquid content, 
although we note that this could also be achieved by changing the liquid 
density or its surface tension.

\subsection{Film in zero gravity}

\label{sect-res-nograv}

\begin{figure}
\begin{center}
\psfig{figure=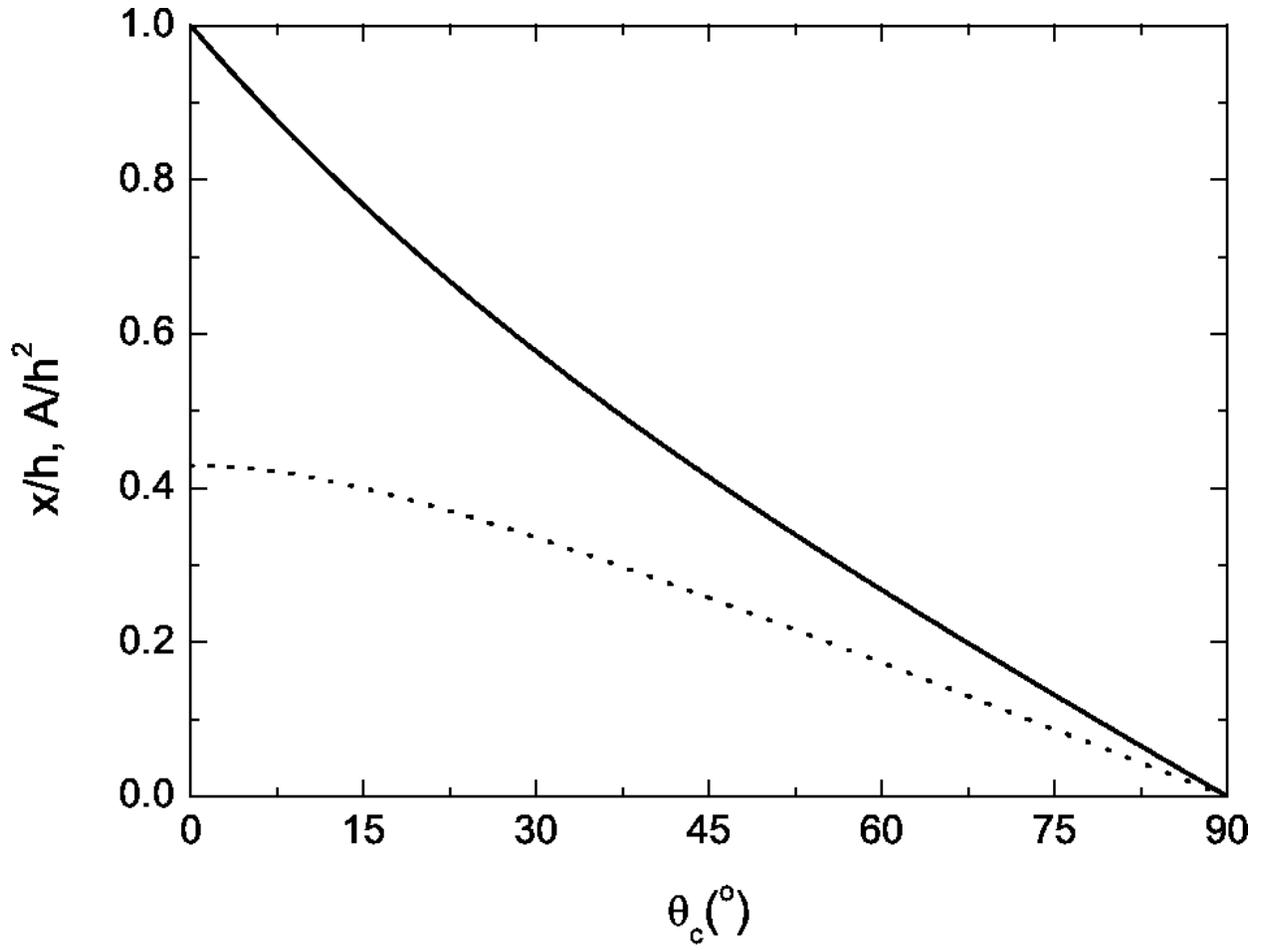,width=\textwidth}
\end{center}
\caption{Dimensionless Plateau border half-width at the substrate 
$x^\prime(z^\prime=0)$ (solid line) and dimensionless Plateau border area 
$A^\prime$ (dotted line) {\it vs} contact angle $\theta_c$  for ${\rm Bo}=0$ 
(corresponding to zero gravity). Recall that in this case the top and bottom 
Plateau borders are identical.}
\label{fig2}
\end{figure}

First, it is instructive to consider the case ${\rm Bo}=0$, corresponding 
to zero gravity, for which the Plateau borders at the top and bottom substrates 
behave identically: their surfaces are arcs of circle. The integrals in 
equations (\ref{defxfinb}) or (\ref{defxfint}), and (\ref{adeffinalb}) or 
(\ref{adeffinalt}) can now be performed analytically, yielding
\begin{eqnarray}
\label{defxfin0}
x^\prime(z^\prime)&=&{1\over{\cos\theta_c}}\left\{1-\left[1-
\left(1-z^\prime\right)^2\cos^2\theta_c\right]^{1/2}\right\}, \\
\label{adeffinal0}
A^\prime&=&{2\over{\cos\theta_c}}\left(1-{1\over 2}\sin\theta_c
-{{{\pi\over 2}-\theta_c}\over{2\cos\theta_c}}\right).
\end{eqnarray}
In particular, the half-width of the Plateau border at the substrate is
\begin{equation}
\label{xpzp0}
x^\prime(z^\prime=0)={{1-\sin\theta_c}\over{\cos\theta_c}}.
\end{equation}
Figure \ref{fig2} plots $A^\prime$ and $x^\prime(z^\prime=0)$ given by 
equations (\ref{adeffinal0}) and (\ref{xpzp0}), respectively. In 
the absence of gravity, the Plateau border can only exist if $\theta_c<\pi/2$,
since its surfaces are circular arcs. In the limit $\theta_c\to 0$, we 
naturally have $x^\prime(z^\prime=0)=1$ and $A^\prime\to 2-\pi/2$, which 
corresponds to twice the difference between the areas of a square of side 
length 1 and of a quarter of a circle of unit radius inscribed in it. 
For $\theta_c\to \pi/2$, on the other hand, both $x^\prime(z^\prime=0)$ 
and $A^\prime$ approach zero, because the film must extend vertically down 
(or up) to meet the substrate.

\subsection{Film in non-zero gravity: bottom Plateau border}

\label{sect-res-bot}

\begin{figure}
\begin{center}
\psfig{figure=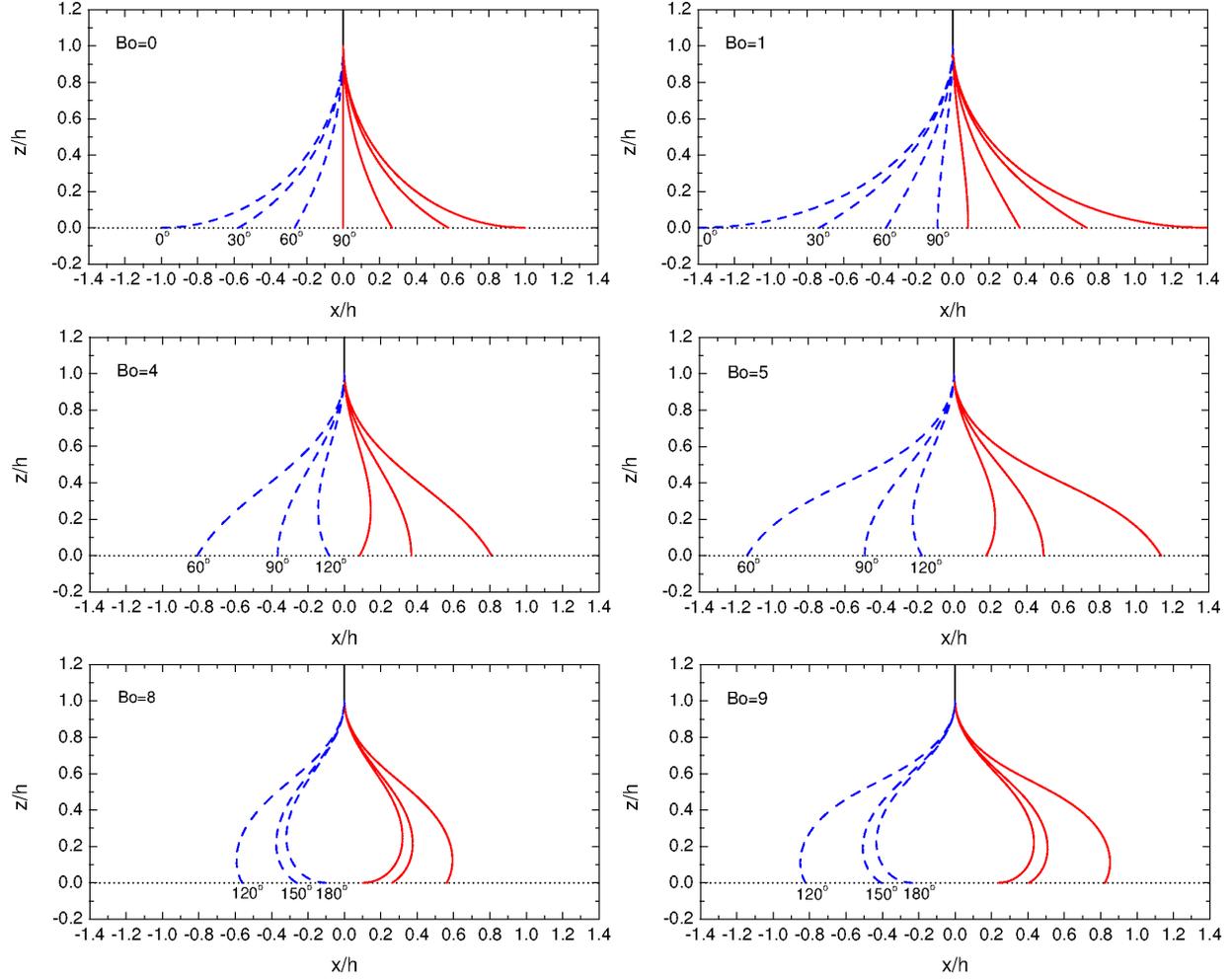,width=\textwidth}
\end{center}
\caption{Analytically-calculated Plateau border shapes at the bottom substrate,
for ${\rm Bo}$ and $\theta_c$ as given. The left-hand air-liquid interfaces 
are shown as dashed blue lines, the right-hand ones as solid red lines.}
\label{fig7}
\end{figure}

\begin{figure}
\begin{center}
\psfig{figure=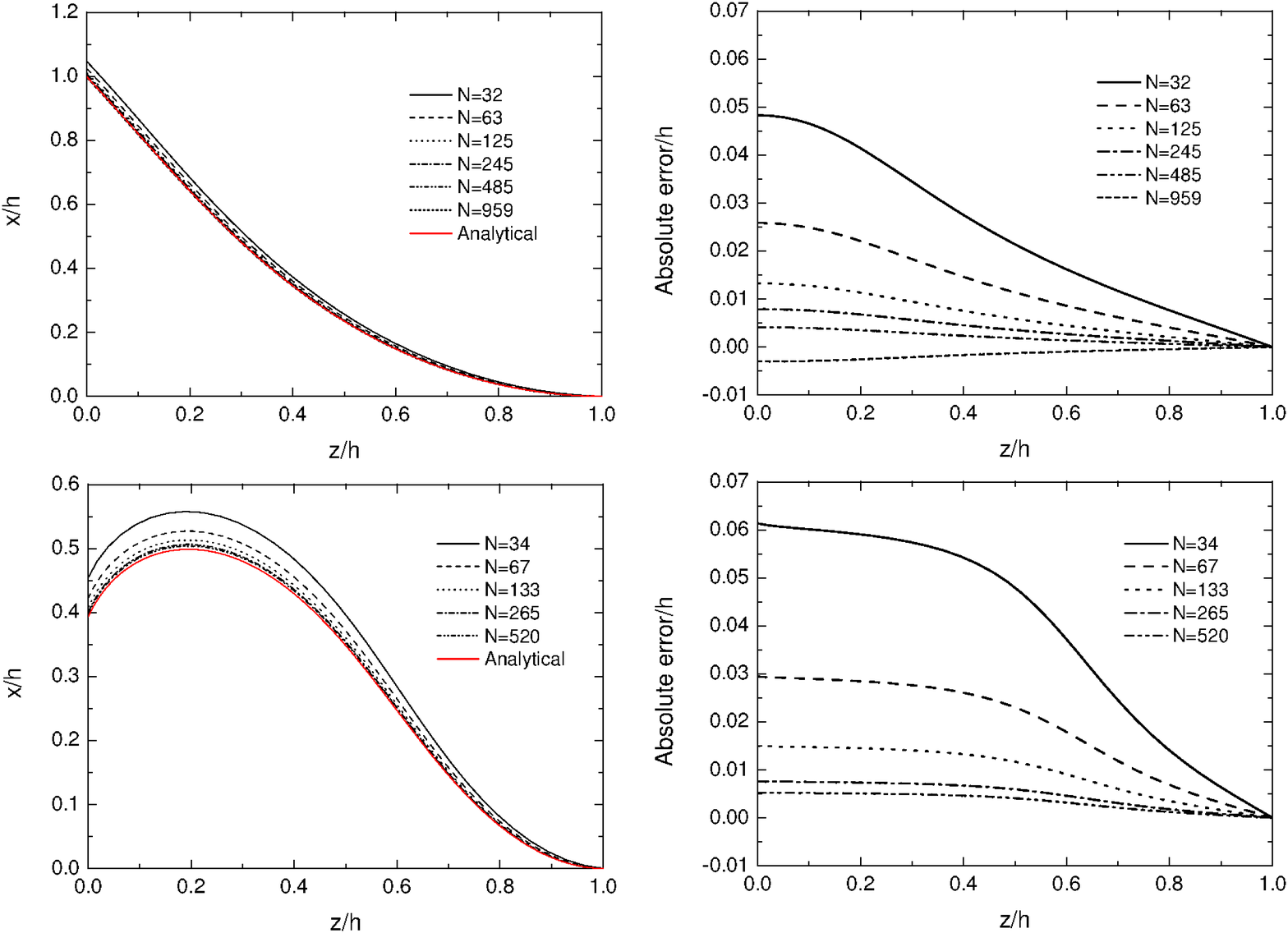,width=\textwidth}
\end{center}
\caption{Left column: Plateau border shapes from analytical theory (red lines) 
and from Surface Evolver with various levels of refinement, as given by the 
number of line segments $N$ used to discretise the interface (black lines). 
Right column: absolute errors at each height $z$, defined as the difference 
between each of the Surface Evolver curves and the analytical theory curve 
in the left panel of the same row. 
Top row: ${\rm Bo}=2.138457$, $\theta_c=30^\circ$. 
Bottom row: ${\rm Bo}=8.975624$, 
$\theta_c=151^\circ$.}
\label{fig8}
\end{figure}

Figure \ref{fig7} displays Plateau border shapes at the bottom substrate 
calculated using equation (\ref{defxfinb}), for various combinations 
of ${\rm Bo}$ and $\theta_c$.
As might be intuitively expected, the Plateau border is widest at the substrate
(i.e., at $z=0$) for $\theta_c<90^\circ$, but above the substrate (i.e., at some
$z=z_{max}>0$) for $\theta_c>90^\circ$. The height $z_{max}$ at which the Plateau
border is widest can be found as a function of $\theta_c$ and ${\rm Bo}$, 
but we do not present it here.

Figure \ref{fig8} compares bottom Plateau border shapes from analytical 
theory and Surface Evolver simulations. Agreement is excellent at small 
contact angles, but less so at larger contact angles and large Bond numbers 
where a very fine discretisation is needed to achieve sufficient accuracy in 
the simulations. This may be due to the considerable range of (concave and 
convex) curvatures of the bounding surfaces that exists in these cases, 
where discretisation errors may tend to accumulate more. In particular, for 
$\theta_c > 90^\circ$, for which $x^\prime(z^\prime)$ has a maximum at $z$
strictly greater than zero, the $x$ location of this maximum is particularly 
sensitive to small errors in the inclination of the interface above it.
In both cases, absolute errors are largest at the substrate ($z=0$),
because the film is pinned at $x=0$ at the Plateau border apex. The other 
main difficulty in the simulations is that of approximating the zero degree 
contact angle at the Plateau border apex with {\it straight} segments; the 
inevitable small error here propagates along the surface, as described above.

Equations (\ref{defxfinb}) and (\ref{adeffinalb}) do not yield physically
meaningful results for all values of $\theta_c$ and ${\rm Bo}$. We next 
discuss the non-trivial conditions defining their domains of validity.

For sufficiently strong gravity (i.e., sufficiently large ${\rm Bo}$), 
the surfaces bounding the Plateau border may become horizontal 
at some point above the substrate, even if they are non-horizontal at the 
substrate, due to the fact that hydrostatic equilibrium favours higher 
pressure (and thus convex curvature) in the lowest parts of these surfaces. 
However, the inclination angle must not change sign, because $x^\prime(z^\prime)$ 
would then become multiple-valued for a single $z^\prime$. This is inconsistent 
with hydrostatic equilibrium, because it would imply a concave curvature 
existing at levels below a convex curvature (as illustrated schematically in 
figure \ref{fig0}c). The Plateau border 
surface may therefore only be horizontal at an inflection point, where 
$d^2 x/dz^2=0$. This situation can be considered a threshold beyond which it 
becomes impossible to satisfy the Young-Laplace law, and thus beyond which the
Plateau border is no longer physically realisable. 
It is therefore essential to determine this threshold. From 
equation (\ref{laplace2d}) it can be seen that the film is horizontal when 
$\Delta p=0$, corresponding to $dx/dz\rightarrow\infty$, or $\cos\theta=1$, 
according to equation (\ref{slopedef}). This produces a singularity in the 
integral in equation (\ref{defxfinb}) when the denominator of the integrand 
vanishes,
\begin{equation}
\frac{\rm Bo}{2}z^{\prime 2}+\left(\cos\theta_c-
\frac{\rm Bo}{2}\right)z^\prime+1 -\cos \theta_c = 0,
\label{2ndorder}
\end{equation}
which gives the vertical coordinates of the points where the film surface is 
horizontal (if they do exist). The solutions to equation (\ref{2ndorder}) are
\begin{equation}
z^\prime=\frac{\frac{\rm Bo}{2}-\cos \theta_c \pm \sqrt{\left( \frac{\rm Bo}{2}
-\cos\theta_c \right)^2 - 2{\rm Bo}( 1- \cos \theta_c )}}{\rm Bo}.
\label{solutz}
\end{equation}
When the expression under the square root is negative, no points exist 
at which the film is horizontal. The threshold where such points
begin to exist occurs when this expression vanishes, namely when
\begin{equation}
{\rm Bo}^2+(4 \cos \theta_c - 8){\rm Bo} + 4 \cos^2 \theta_c = 0.
\label{2ndobo}
\end{equation}
The solutions to this second-order algebraic equation for ${\rm Bo}$ are
\begin{equation}
{\rm Bo} =  4 - 2 \cos \theta_c \pm \sqrt{( 4 - 2 \cos \theta_c )^2 - 
4 \cos^2 \theta_c}.
\label{solutbo}
\end{equation}

It is clear from equation (\ref{solutz}) that the situation where two real 
solutions for $z^\prime$ exist, i.e., the domain of parameter space that is 
unphysical, corresponds to values of ${\rm Bo}$ outside the interval 
bounded by the solutions given by equation (\ref{solutbo}). It can be shown, 
however, that in the lower range of values outside this interval, 
$\cos\theta=1$ is not fulfilled for $0 < z^\prime <1$, so only the inequality 
involving the largest root (with the plus sign in equation (\ref{solutbo})) 
is relevant; the model then becomes unphysical when
\begin{equation}
{\rm Bo} >  4 - 2 \cos \theta_c + \sqrt{( 4 - 2 \cos \theta_c )^2 - 
4 \cos^2 \theta_c}.
\label{invalid}
\end{equation}
For each value of $\theta_c$, the right-hand side of equation (\ref{invalid}) 
defines an upper bound for ${\rm Bo}$, or equivalently an upper bound for $h$, 
for which the bottom Plateau border is physically realisable and the surface
foam-philic.

Another condition for the validity of equations (\ref{defxfinb}) and 
(\ref{adeffinalb}) follows from requiring that the Plateau border is 
topologically sound. Equation (\ref{defxfinb}) specifies the horizontal 
coordinate of the right-hand surface at the substrate, $x^\prime(z^\prime=0)$.
In the most usual situations, the contact angle $\theta_c$ lies between 
$0$ and $\pi/2$ (hydrophilic surface), which implies that $x^\prime(z^\prime)$ 
given by equation (\ref{defxfinb}) is always positive, as $z^\prime<1$ by 
definition. When the substrate is hydrophobic ($\pi/2 < \theta_c <\pi$), 
however, the numerator of the fraction in the integrand of equation 
(\ref{defxfinb}) may become negative, and therefore $x^\prime(z)$ may also be 
negative. This, which is easiest to fulfil for $x^\prime(z^\prime=0)$ (as 
the term involving ${\rm Bo}$ in the numerator is always non-negative), 
is unphysical, since $x^\prime(z^\prime=0)<0$ would correspond to Plateau border
surfaces that cross each other before reaching the substrate (see figure 
\ref{fig0}b).

Equations (\ref{defxfinb}) and (\ref{adeffinalb}) are therefore only valid 
outside the interval defined by equation (\ref{invalid}) and when
$x^\prime(z^\prime=0)\ge 0$. the Plateau border area 
$A^\prime$ given by equation (\ref{adeffinalb}) 
may also be negative for hydrophobic substrates ($\theta_c>\pi/2$), but the 
domain of parameter space where $A^\prime<0$ is contained in the (equally 
unphysical) domain where $x^\prime(z^\prime=0)<0$, because one may have 
$x^\prime(z^\prime=0)<0$, but $x^\prime(z^\prime>0)>0$. That is, the (negative) 
area below the point where the Plateau border surfaces cross may not fully 
compensate for the (positive) area above that point. Although the domains 
defined by $x^\prime(z^\prime=0)<0$ or $A^\prime<0$ are unphysical, one can still 
solve the Young-Laplace equation and find $x^\prime(z^\prime)$ and $A^\prime$ in 
such cases. Note that, on the contrary, this is not possible in the forbidden 
domain defined by equation (\ref{invalid}), because in that case the 
Young-Laplace equation breaks down.

The above findings are summarised in figure \ref{fig1}a.
The cross-hatched domain is where inequality (\ref{invalid}) 
is satisfied and hence where there can be no Plateau border because 
no solution to the Young-Laplace equation exists. The shaded domain 
is where $x^\prime(z^\prime=0)<0$, i.e., the left and right Plateau 
border surfaces intersect before meeting the substrate or switch places 
altogether. Examples of Plateau border shapes in
this domain are given in figure \ref{fig5} (top row). Both 
cross-hatched and shaded domains thus consist of $(\theta_c,{\rm Bo})$ pairs 
for which no bottom Plateau border can exist -- `forbidden' states -- separated 
by a white band of `allowed' states. Furthermore, allowed Plateau borders may
exhibit an inflection point, at which the curvature of their liquid-vapour 
interfaces changes from convex near the substrate to concave near the apex. 
Since inflection points correspond to $\Delta p=0$, they will first appear 
when this condition is fulfilled at the substrate, $z=0$. From equation 
(\ref{defpres}) and using the definition of ${\rm Bo}$, we obtain the threshold
\begin{equation}
{\rm Bo}=2\cos \theta_c,
\label{ipthresh}
\end{equation}
which is plotted as the dashed line in figure \ref{fig1}a. Below this line, 
Plateau borders do not have inflection points; above it they do, owing 
to the effect of gravity. The $z$ coordinate of the inflection point 
can also be found from the theory, but this is beyond the scope of 
the present study. Clearly, most realisable Plateau borders do have 
inflection points, i.e., the curvature of their surfaces changes 
sign, from convex near the substrate to concave nearer the apex. 

The solid curves inside the white (allowed) and 
shaded (forbidden) parameter domains are lines of constant $x^\prime(z^\prime=0)$ 
as labelled. At constant $\theta_c$, $x^\prime(z^\prime=0)$ increases as 
${\rm Bo}$ is increased, which seems an intuitive effect of gravity. 
The same qualitative trend occurs for $A^\prime$ (not shown). 
In the white domain $x^\prime(z^\prime=0)$ varies from 0 at the lower 
boundary to a value that is a function of $\theta_c$, but always greater than 
1, at the upper boundary. This latter limit corresponds to a situation where 
the Plateau border is strongly `flattened' by gravity. We could not determine
any bound for $x^\prime(z^\prime=0)$ (or for the corresponding area $A^\prime$) as 
${\rm Bo}$ approaches this limit. This means that both quantities can
potentially become very large, although the range of ${\rm Bo}$ in which 
this occurs is very narrow, and therefore should be difficult to access in
practice.

Figure \ref{figmenisci} shows photographs of Plateau borders (equivalent of 
side view in figure \ref{figdrops}b) at the liquid film-surface interface for 
four of the five surfaces used in the experiments, overlaid with their
analytically-calculated shapes for the same Bond numbers and contact angles.
Then figure \ref{figcompare} compares theoretical predictions and experimental 
results for the Plateau border half-width $x$, scaled by its height $h$, 
{\it vs} Bond number. (Equation (\ref{defxfinb}) has been 
used in both cases.) The general trends of $x/h$ are well reproduced, with 
the only substantial deviation occurring for the most hydrophobic substrate 
(teflonised black silicon) at ${\rm Bo}\approx 8$. Since the vertical 
asymptotes of the theoretical curves correspond to the upper ${\rm Bo}$ limit 
mentioned in the preceding paragraph, it is to be expected that experimental 
results in these regions should be more sensitive to, for example, errors in 
measuring $h$, from which ${\rm Bo}$ is calculated. This might explain the 
poorer agreement between theory and experiment in the upper ${\rm Bo}$ range 
of each curve. One other possible source of discrepancy is 
contact angle hysteresis, which is neglected in our theory and simulations
but should be more pronounced at large ${\rm Bo}$.

\subsection{Film in non-zero gravity: top Plateau border}

\label{sect-res-top}

A similar analysis can be performed to determine the validity of equations
(\ref{defxfint}) and (\ref{adeffinalt}) for the top Plateau border. 
Note that now the $z^\prime$-axis is directed downwards from $z^\prime=0$ (the 
top substrate). Since, because of hydrostatic equilibrium, their curvature 
must become less convex, or more concave, as $z^\prime$ decreases, the only way 
that the surfaces can become horizontal at an inflection point before reaching 
the substrate (which defines a threshold for the existence of solutions
to the Young-Laplace equation) is by having convex curvature at the bottom. 
This requires that the film surfaces cross (unphysically) immediately at the 
apex where the Plateau border meets the planar film underneath.
The condition to be fulfilled for the existence of a solution to the Plateau 
border surfaces is then $\cos\theta=-1$, which corresponds to a singularity 
in the integrand of equation (\ref{defxfint}) if
\begin{equation}
\frac{\rm Bo}{2}z^{\prime 2}-\left(\cos\theta_c+
\frac{\rm Bo}{2}\right)z^\prime+1 +\cos \theta_c = 0,
\label{2ndorder2}
\end{equation}
yielding
\begin{equation}
z^\prime=\frac{\frac{\rm Bo}{2}+\cos \theta_c \pm \sqrt{\left( \frac{\rm Bo}{2}
+\cos\theta_c \right)^2 - 2{\rm Bo}( 1+ \cos \theta_c )}}{\rm Bo}.
\label{solutz2}
\end{equation}
Now, the condition for the threshold at which $x^\prime(z^\prime)$ becomes 
multivalued is
\begin{equation}
{\rm Bo}^2-(4 \cos \theta_c + 8){\rm Bo} + 4 \cos^2 \theta_c = 0,
\label{2ndobo2}
\end{equation}
the solution of which is
\begin{equation}
{\rm Bo}= 4+2\cos\theta_c \pm \sqrt{(4+2\cos\theta_c )^2-4\cos^2 \theta_c}.
\label{solutbo2}
\end{equation}
A similar argument as used previously applies to the two roots of equation  
(\ref{solutbo2}), so that we take the largest root to allow us to predict
the `forbidden' domain of parameter space where the Young-Laplace equation 
has no solution:
\begin{equation}
{\rm Bo} >  4 + 2 \cos \theta_c + \sqrt{( 4 + 2 \cos \theta_c )^2 - 
4 \cos^2 \theta_c}.
\label{invalid2}
\end{equation}
This equation is equivalent to equation (\ref{invalid}) if the sign of 
$\cos \theta_c$ is reversed.

Note that, as with the bottom Plateau border, in the domain of parameter 
space where inequality (\ref{invalid2}) is not satisfied there are many 
$(\theta_c,{\rm Bo})$ pairs for which the Young-Laplace equation has a 
solution, but $x^\prime(z^\prime=0)<0$ and $A^\prime<0$, which is obviously 
unphysical on topological grounds.
However, neither of these criteria may now be used to delimit 
the allowed domains of parameter space, as there are solutions with  
$x^\prime(z^\prime=0)>0$ or $A^\prime>0$ for which the two Plateau border 
surfaces still cross. Since, from hydrostatic equilibrium, the most convex 
curvature of the Plateau border surfaces must exist near their lowest 
point, this is where they are most likely to cross. The only way to avoid 
this topological violation is by requiring that the curvature should not 
be convex at the point where the Plateau border surfaces meet the planar 
film below. Hence, the threshold condition for the realisability of the 
Plateau border is, in this case, having zero curvature at the lower end of the 
surfaces bounding the Plateau border, i.e., $d^2 x/dz^2(z=h)=0$, thereby 
avoiding convex curvature altogether. This condition, again, corresponds to 
$\Delta p=0$. Given the definition of $\Delta p$ for the top Plateau border, 
namely
\begin{equation}
\Delta p=p_{b0}-p_a +\rho g z, 
\label{prestop}
\end{equation}
and the modified form of equation (\ref{defpres}) that results,
\begin{equation}
\frac{p_a-p_{b0}}{\gamma}=\frac{1}{h}\cos\theta_c+\frac{\rho g h}{2 \gamma},
\label{defpres2}
\end{equation}
equation (\ref{prestop}) can be inserted into equation (\ref{defpres2}) 
for $z=h$ and $\Delta p=0$ to yield
\begin{equation}
{\rm Bo} = 2\cos \theta_c.
\label{dashed}
\end{equation}
Interestingly, this is exactly the same as the threshold for a bottom Plateau 
border to have an inflection point, equation (\ref{ipthresh}). The difference 
here is that, since (by the above arguments) a top Plateau border cannot have
any inflection points, equation (\ref{dashed}) now assumes the much more 
important role of defining an upper bound for ${\rm Bo}$ beyond which no top 
Plateau border can exist.

\begin{figure}
\begin{center}

(a)
\vspace{0.25cm}

\psfig{figure=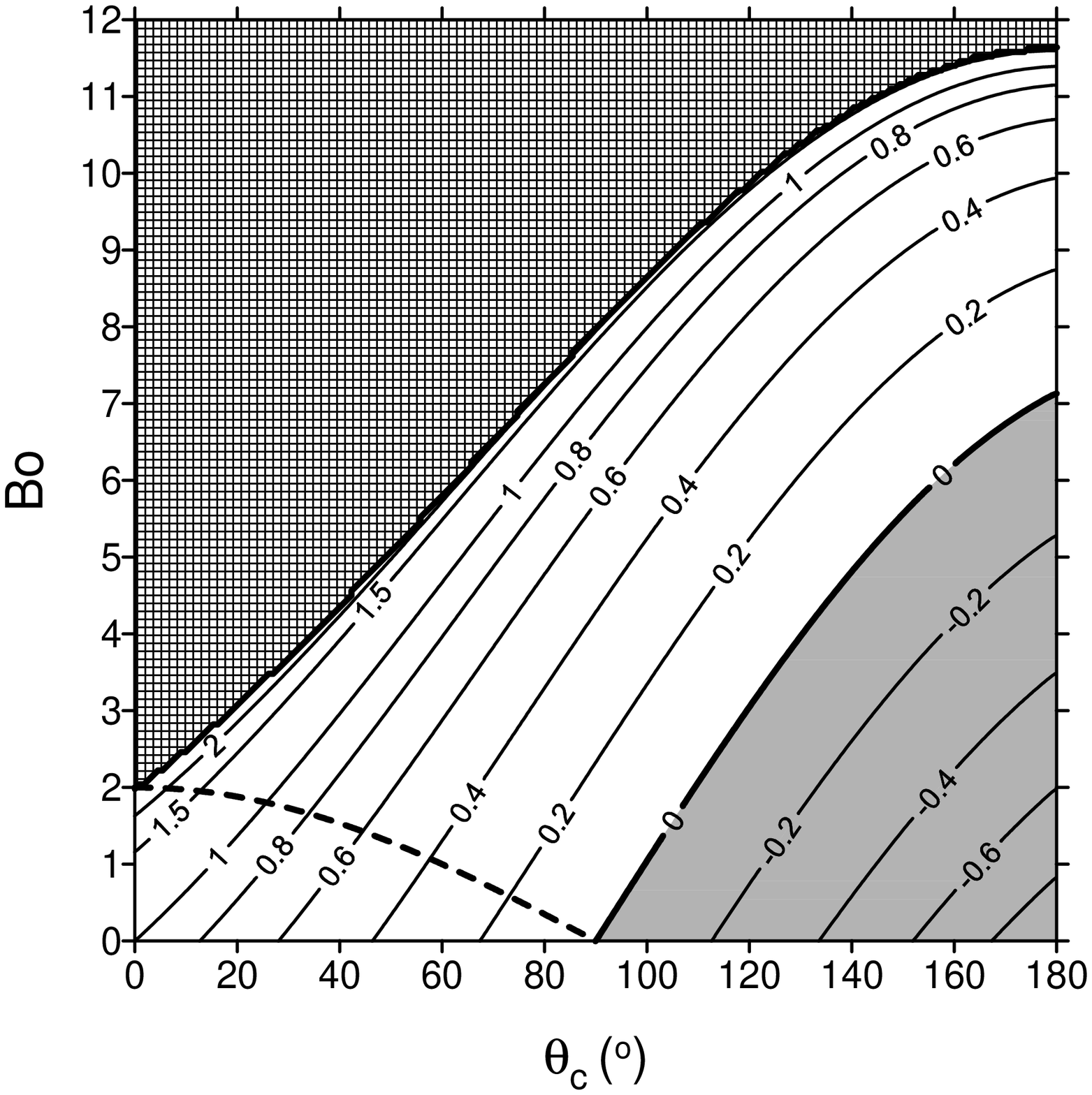,width=9cm}

(b)
\vspace{0.25cm}

\psfig{figure=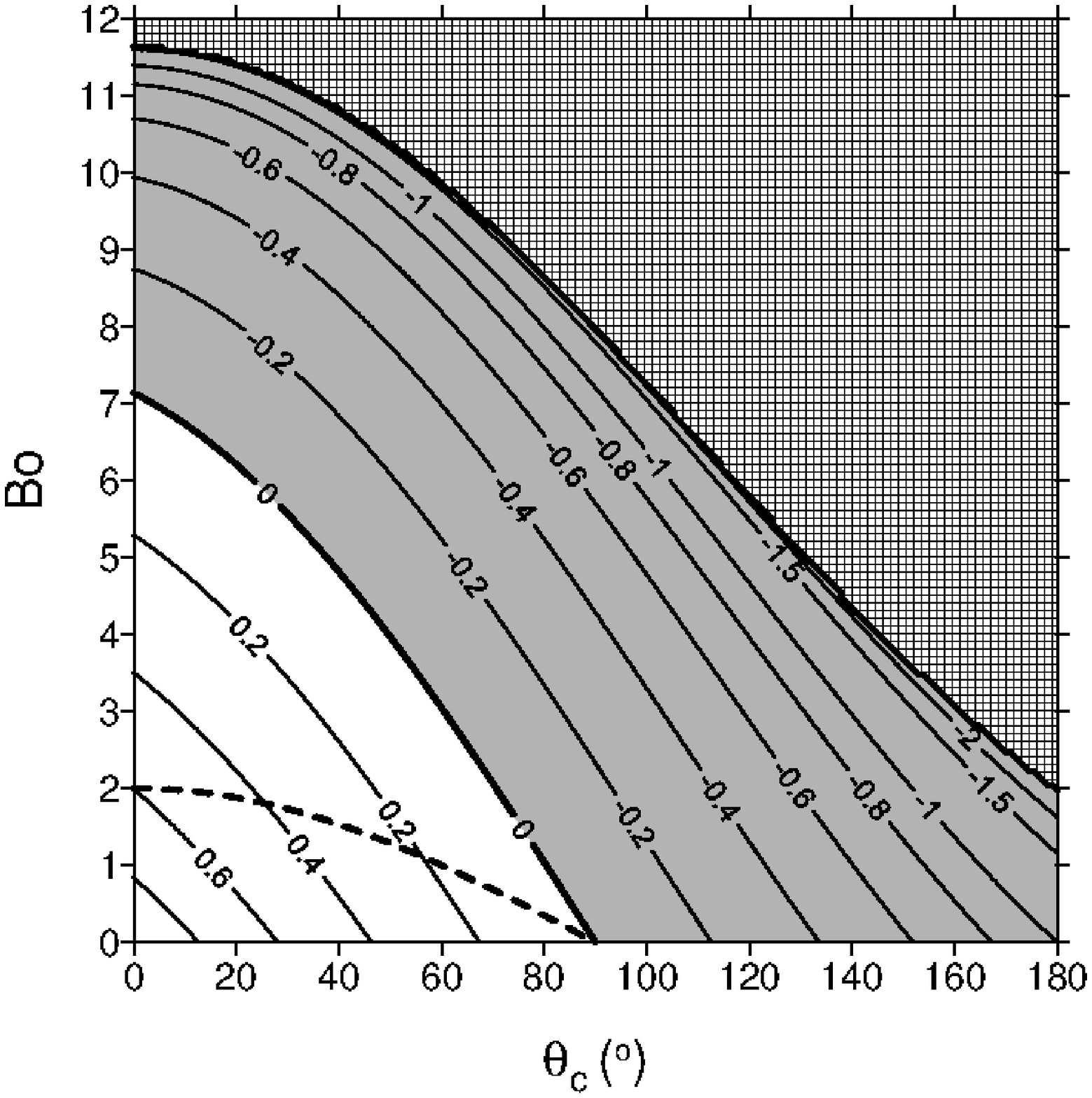,width=9cm}
\end{center}
\caption{Domains of allowed and forbidden Plateau borders at (a) the bottom 
substrate, and (b) the top substrate, in the space of liquid contact 
angle $\theta_c$ and Bond number ${\rm Bo}$. The curves are lines of 
constant $x^\prime(z^\prime=0)$ as labelled. See the text for details.}
\label{fig1}
\end{figure}

\begin{figure}
\begin{center}
\psfig{figure=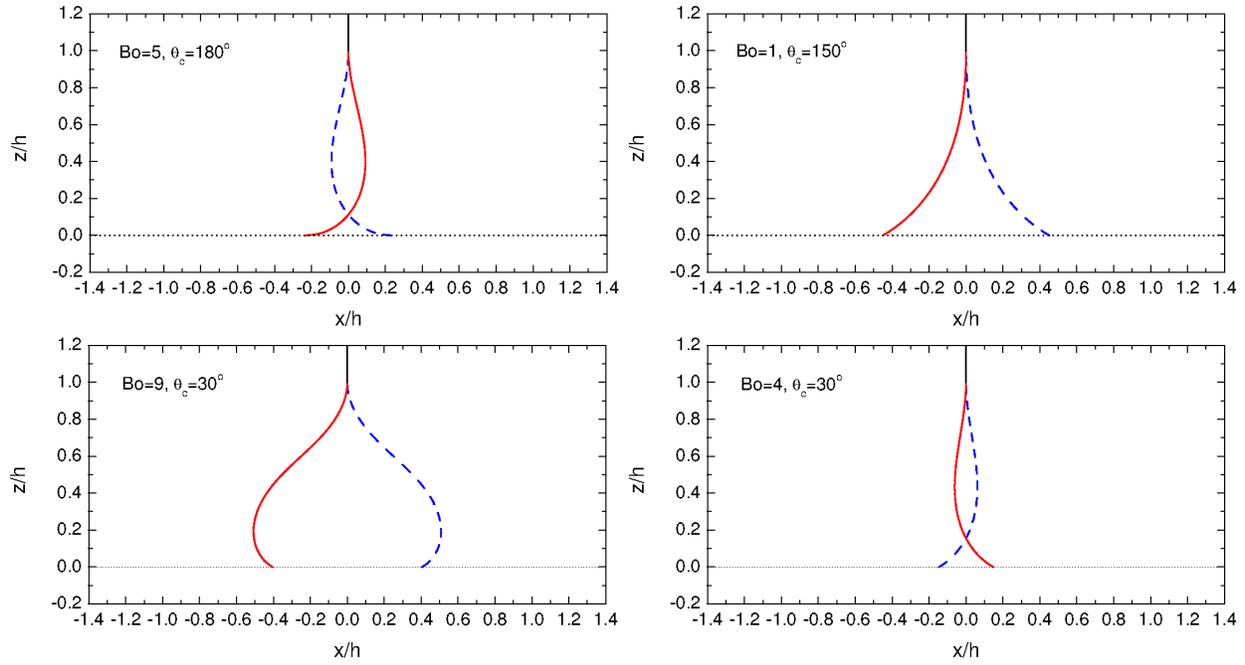,width=\textwidth}
\end{center}

\caption{Examples of unphysical Plateau borders in the 
shaded domain of figure \ref{fig1}a (top row); the shaded domain of 
figure \ref{fig1}b (bottom left); and the white domain of figure \ref{fig1}b, 
above the dashed line (bottom right)}
\label{fig5}
\end{figure}

\begin{figure}
\begin{center}
\psfig{figure=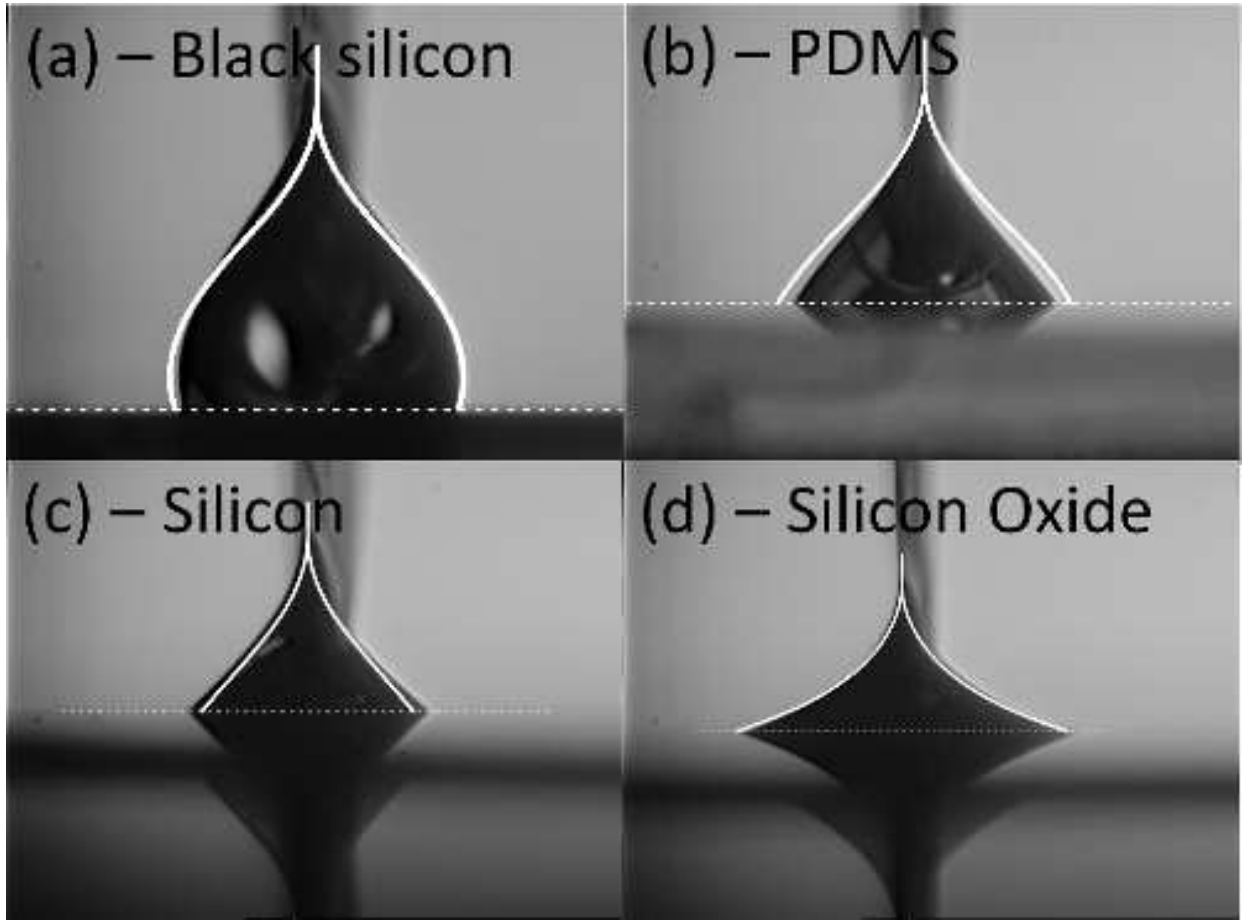,width=\textwidth}
\end{center}

\caption{Plateau borders at the liquid film-surface interface for four of 
the five substrates used in the experiments. (a) Teflonised black silicon; 
(b) PDMS elastomer; (c) teflonised polished silicon; and (d) silicon oxide.
The Bond numbers and Plateau border base widths are: (a) ${\rm Bo}=6.65$ 
and 3.8~mm; (b) ${\rm Bo}=3.36$ and 3.7~mm; (c) ${\rm Bo}=2.13$ and 3.4~mm;
and (d) ${\rm Bo}=1.59$ and 4.7~mm. The solid white lines are the 
analytically-calculated Plateau border shapes for the same Bond number 
and contact angle.}
\label{figmenisci}
\end{figure}

\begin{figure}
\begin{center}
\psfig{figure=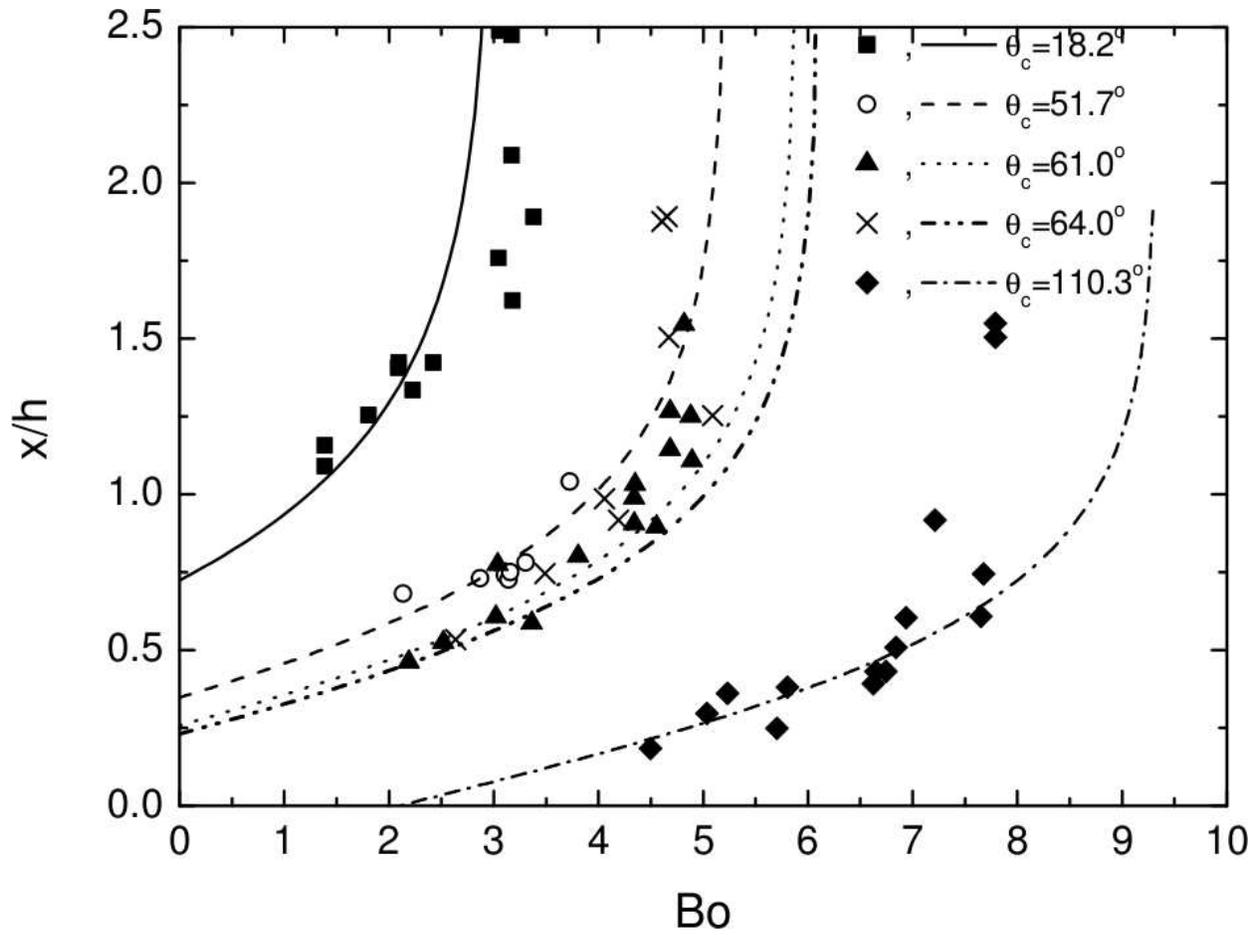,width=\textwidth}
\end{center}
\caption{Scaled Plateau border half-width $x/h$ {\it vs} Bond number, for all
five substrates investigated. The curves are theoretical predictions, symbols 
are experimental data points. }
\label{figcompare}
\end{figure}

The above findings are summarised in figure \ref{fig1}b. As in figure 
\ref{fig1}a, the white domain comprises $(\theta_c,{\rm Bo})$ pairs for 
which the Plateau border half-width at the (in this case top) substrate 
is positive; as explained above, this is a necessary (but not 
sufficient) condition for the Plateau border to be physically realisable. 
In the shaded domain, by contrast, $x^\prime(z^\prime=0)<0$.

Although, as for the bottom Plateau border, equation (\ref{defxfint}) can 
still be solved in the shaded domain of figure \ref{fig1}b, the resulting 
Plateau borders are unphysical. In the cross-hatched 
domain, which is a mirror image of that found for the bottom Plateau border, 
the Young-Laplace equation has no solution. However, in contrast to figure 
\ref{fig1}a, the white region in figure \ref{fig1}b does not now coincide with
the domain where the Plateau border is realisable: this is only so in the much
smaller domain below the dashed line, which is given by equation 
(\ref{dashed}). In other words, at the top substrate only Plateau borders 
with no inflection points can exist  -- their surfaces are always concave.
Examples of unphysical top Plateau border shapes are provided in figure 
\ref{fig5} (bottom row). Note also that both $x^\prime(z^\prime=0)$ (shown 
in figure \ref{fig1}b) and $A^\prime$ (not shown) decrease as ${\rm Bo}$ 
is increased at constant $\theta_c$, which again is expected given the 
direction of gravity.

\begin{figure}
\begin{center}
\psfig{figure=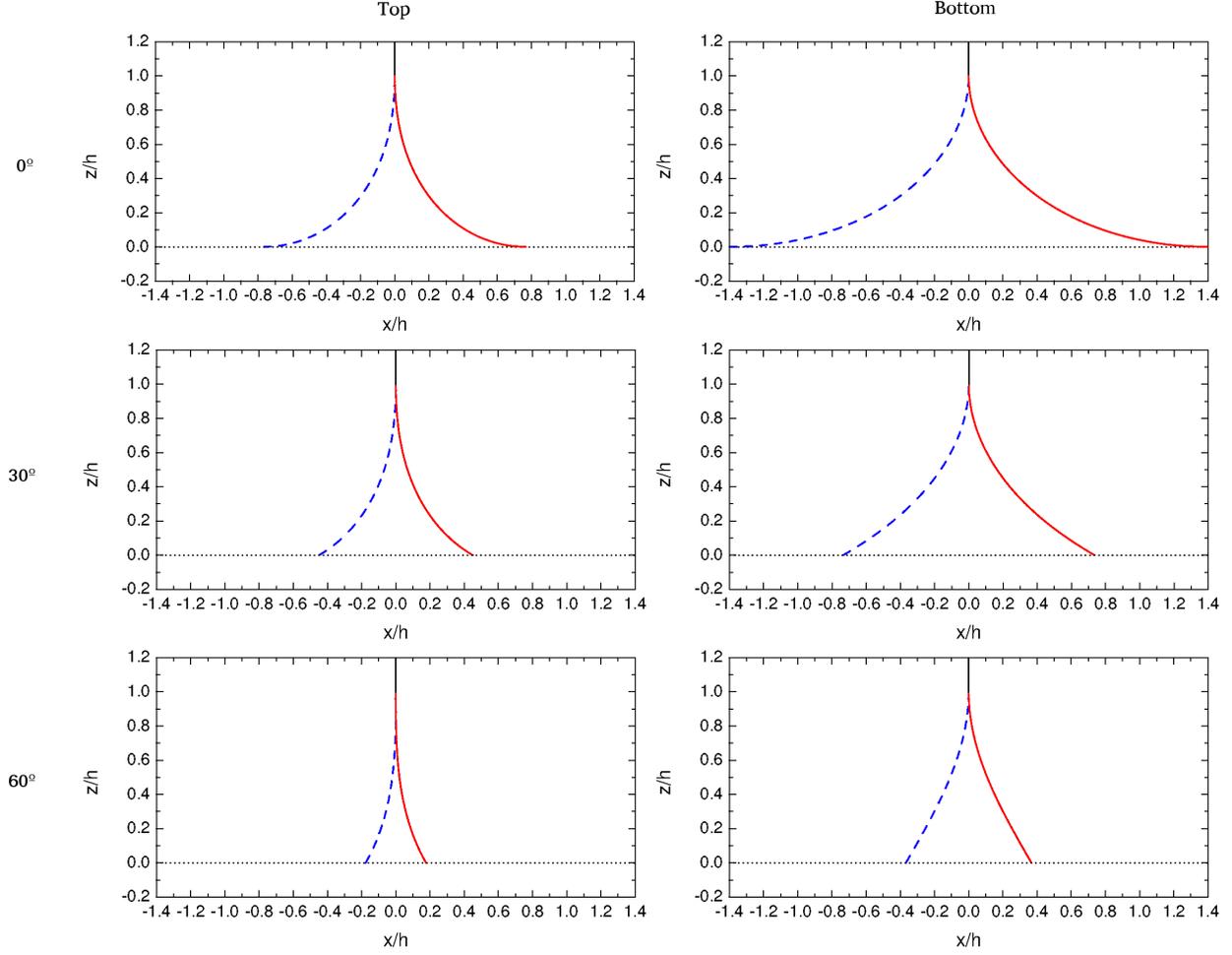,width=\textwidth}
\end{center}
\caption{Analytically-calculated Plateau border shapes at the top (left column) 
and bottom (right column) substrates, for ${\rm Bo}=1$ and $\theta_c=0^\circ$ 
(top row), $30^\circ$ (centre row) and $60^\circ$ (bottom row). The left-hand 
air-liquid interfaces are shown as dashed blue lines, the right-hand ones as 
solid red lines. The Plateau borders at the top substrate are shown inverted 
for ease of comparison.}
\label{fig6}
\end{figure}

As might be intuitively expected, Plateau borders can only exist at the top 
substrate if the liquid contact angle $\theta_c\le\pi/2$, otherwise the liquid
will just detach from the substrate. Values of $x^\prime(z^\prime=0)$ in the white
domain below the dashed line in figure \ref{fig1}b are all below 1, which 
illustrates how gravity acts to stretch the top Plateau border vertically 
(and consequently compress it horizontally), especially for the largest 
allowed values of ${\rm Bo}$, as can be seen in figure \ref{fig6}
(calculated using equations (\ref{defxfinb}) and 
(\ref{defxfint}) for the bottom and top Plateau borders, respectively). 

\begin{figure}
\begin{center}
\psfig{figure=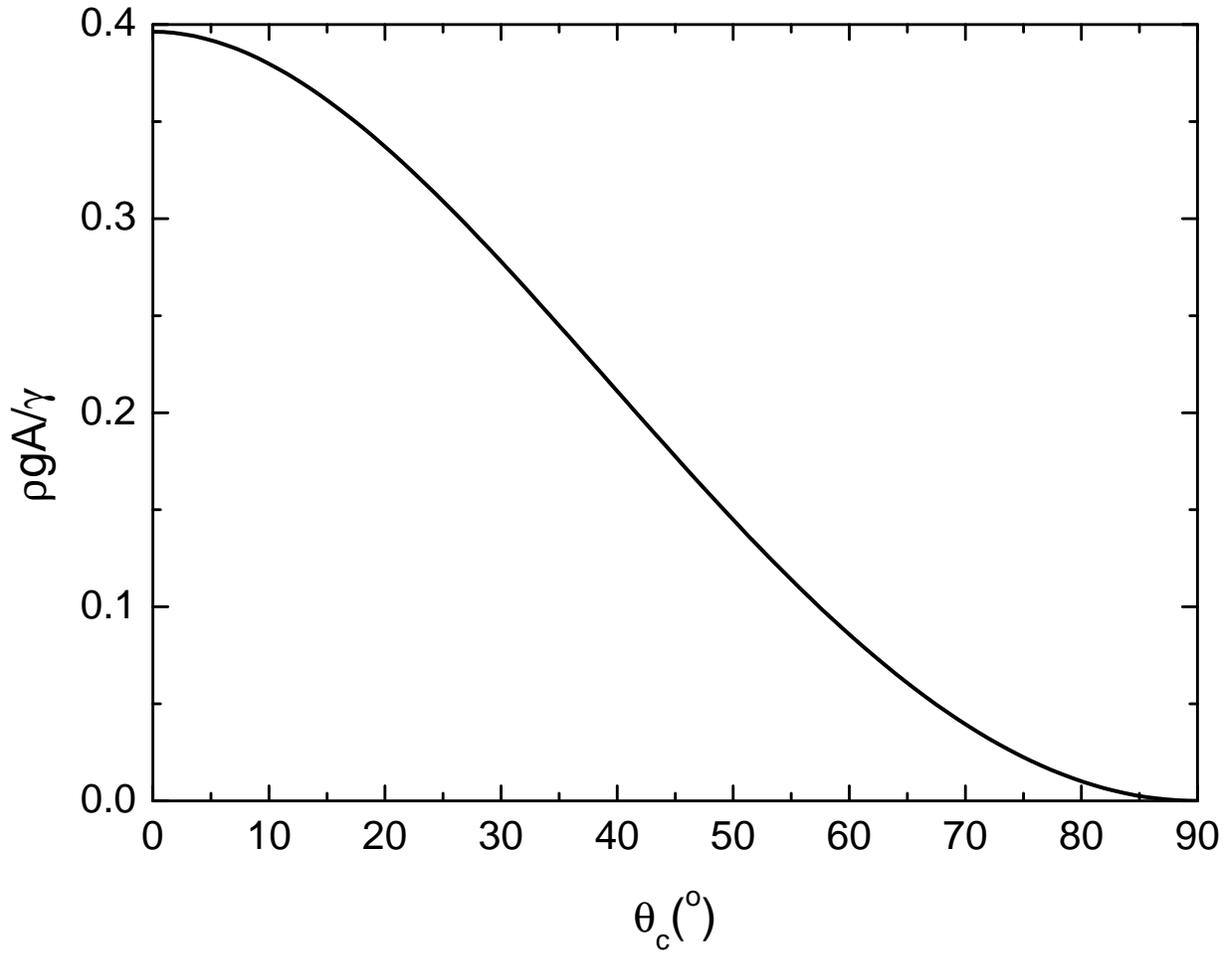,width=\textwidth}
\end{center}
\caption{Maximum top Plateau border area, normalised by square of capillary 
length, {\it vs} contact angle $\theta_c$.}
\label{fig9}
\end{figure}

A relevant question that may be asked is: how large, 
in physical dimensions, can the Plateau borders be? 
Given our comments above, about $x^\prime$ and $A^\prime$ being 
unbounded as ${\rm Bo}$ approaches its upper limit, the bottom Plateau border 
can probably be indefinitely large, expanding laterally as more fluid is added 
to it. On the other hand, the answer for the top Plateau border is totally 
different. First, as noted above, no top Plateau border can exist on a 
hydrophobic substrate ($\theta_c>90^\circ$), since it would detach due to 
gravity. When the substrate is hydrophilic ($\theta_c<90^\circ$), however, there 
is an upper bound to the size of the top Plateau border, which depends on the 
contact angle, and naturally approaches zero as $\theta_c\rightarrow 90^\circ$. 
The area of the top Plateau border given by equation (\ref{adeffinalt}) is 
normalised by $h^2$, so it does not give us information about the physical 
size of the Plateau border. A more useful quantity is obtained by multiplying 
$A^\prime$ by ${\rm Bo}$, which gives $\rho g A/\gamma\equiv A/\lambda_c^2$, 
i.e., the Plateau border area normalised by the square of the capillary length 
$\lambda_c=\left(\gamma/\rho g\right)^{1/2}$. Whereas for ${\rm Bo}$ in the 
range $(0,2\cos \theta_c)$ $A^\prime$ attains maximum values
for ${\rm Bo}=0$ (and an absolute maximum for $\theta_c=0$), 
$\rho g A/\gamma$ attains its maximum values for 
${\rm Bo}=2\cos \theta_c$. Figure \ref{fig9} shows how the maximum of 
$A/\lambda_c^2$ (calculated using equation (\ref{adeffinalt})
for ${\rm Bo}=2 \cos \theta_c$) varies as a function 
of $\theta_c$. It can be seen that $A/\lambda_c^2$ attains an absolute maximum 
of 0.396 for $\theta_c=0$. Not surprisingly, this indicates that this maximum 
of $A$ is of the order of the capillary length squared. Using the experimental 
values $g=9.81~{\rm ms}^{-2}$ and $\gamma=28~{\rm mN/m}$ yields an 
absolute maximum for $A$ of $1.138~{\rm mm}^2$.

\section{Conclusions}

\label{sect-disc}

We have studied the shapes of the Plateau borders at which a vertical 
planar liquid film meets horizontal substrates of various wettabilities, 
by analytical theory, numerical simulation, and experiment. The overall 
picture that emerges is that the Plateau borders, and consequently the film
to which they are attached, spanning the gap between the two substrates, can 
only be realised in certain ranges of Plateau border sizes, which are in
turn functions of the liquid contact angle. In other words, a foam-surface 
system can be either `foam-phobic' or `foam-philic'. The Plateau border at the 
top substrate has quite a small domain of existence and a necessary condition 
is that the liquid contact angle is less than $90^\circ$. Its maximum area 
decreases as the contact angle increases, and attains an absolute maximum 
of $0.396$ times the square of the capillary length, for $\theta_c=0$. 
The  Plateau border at the bottom substrate has a larger domain of existence, 
larger contact angles being required  at higher Bond numbers and vice versa. 
The practical importance of this is that both surface and liquid (foam) 
properties need to be taken into account in applications where wetting of 
surfaces by foams plays a role. It suggests, e.g., that self-cleaning surfaces 
for foams could be designed and built.

We are currently working on generalising our results to a bubble on a solid
substrate. We expect qualitatively the same results, although the detailed
shapes of the `allowed'  and `forbidden' domains in $(\theta_c,{\rm Bo})$ 
parameter space will likely be different.

\section*{Acknowledgements}

The work of S. A. was partly supported by the French RENATECH network. 
S. J. C. thanks K. Brakke for his development and maintenance of the Surface 
Evolver code and acknowledges funding from the MSCA-RISE project Matrixassay 
(ID: 644175). P. I. C. T. acknowledges financial support from the 
Funda\c{c}\~{a}o para a Ci\^{e}ncia e Tecnologia (Portugal) through 
contracts nos.\ EXCL/FIS-NAN/0083/2012 and UID/FIS/00618/2013. 
S. A. would very much like to thank Thomas Arscott for fruitful 
discussions concerning the experimental setup. We are grateful 
to W. Drenckhan for a critical reading of the manuscript.

\newpage

\begin{table}[ht]
\begin{minipage}{\textwidth}
\begin{center}
\begin{tabular}{|c|c|c|}
\hline
                 &Material   &Contact angle to bubble solution (deg) \\
\hline
Surface 1        &Silicon oxide                       &$18.2\pm 2.8$ \\
\hline
Surface 2        &Teflonised polished silicon         &$51.7\pm 0.3$ \\
\hline
Surface 3        &PDMS elastomer                      &$61.0\pm 2.1$ \\
\hline
Surface 4        &Teflonised rough silicon            &$64.0\pm 0.4$ \\
\hline
Surface 5        &Teflonised black silicon            &$109.3\pm 0.3$ \\
\hline
\end{tabular}
\end{center}
\caption{List of surfaces prepared and used in this study and their measured 
wetting contact angle with the commercial bubble solution.}
\label{table1}
\end{minipage} 
\end{table}

\newpage

\newpage

\newpage

\newpage

\newpage

\newpage

\newpage

\newpage

\newpage

\newpage

\newpage

\newpage

\newpage

\newpage

\end{document}